\documentclass[12pt]{article}
\usepackage[margin=1.05in]{geometry}
\usepackage[utf8]{inputenc}
\usepackage{graphicx}
\usepackage{setspace}%\usepackage{dcolumn}
\usepackage{bm}
\usepackage{caption}
\usepackage{subcaption}
\usepackage[backend=biber,
            style=phys,
            sorting=none]{biblatex}
\usepackage[linguistics]{forest}
\usepackage{booktabs}
\usepackage[version=4]{mhchem}
\usepackage{color, colortbl}
\usepackage{amsmath}
\usepackage{physics}
\usepackage{setspace}
\usepackage{soul}
\usepackage{authblk}
\usepackage{color,soul}
\usepackage{threeparttable}
\usepackage{float}
\usepackage{import}
\usepackage{placeins}
\usepackage{hyperref}

\usepackage{booktabs, colortbl, xcolor, siunitx, multirow}
\usepackage{adjustbox}
\usepackage{lineno}

\begin{document}
\onehalfspacing
\title{Discovery of a new weberite-type antiferroelectric: La$_3$NbO$_7$ }

\author[1]{Louis Alaerts}
\author[2,3]{Jesse Schimpf}
\author[3,4]{Xinyan Li}
\author[1]{Jiongzhi Zheng}
\author[5,6]{Ella Banyas}
\author[5,6,7]{Jeffrey B. Neaton}
\author[6,8]{Sinéad M. Griffin}
\author[3,4,9]{Yimo Han}
\author[3,4,6,10]{Lane W. Martin}
\author[1,3,4,*]{Geoffroy Hautier}

\affil[1]{Thayer School of Engineering, Dartmouth College, Hanover, NH 03755, USA}
\affil[2]{Department of Materials Science and Engineering, University of California, Berkeley, Berkeley, CA 94720, USA}
\affil[3]{Rice Advanced Materials Institute, Rice University, Houston, TX 77005, USA}
\affil[4]{Department of Materials Science and NanoEngineering, Rice University, Houston, TX 77005, USA}
\affil [5]{Department of Physics, University of California, Berkeley, Berkeley, CA 94720, USA}
\affil[6]{Materials Sciences Division, Lawrence Berkeley National Laboratory, Berkeley, Berkeley, CA 94720, USA}
\affil[7]{Kavli Energy Nanosciences Institute, University of California, Berkeley, Berkeley, CA 94720, USA}
\affil[8]{Molecular Foundry, Lawrence Berkeley National Laboratory, Berkeley, Berkeley, CA 94720, USA}
\affil[9]{Smalley-Curl Institute, Rice University, Houston, TX 77005, USA}
\affil[10]{Departments of Chemistry and Physics and Astronomy, Rice University, Houston, TX 77005, USA}
\affil[*]{Corresponding author: Geoffroy Hautier, geoffroy.hautier@rice.edu}

\date{\today}% It is always \today, today,
             %  but any date may be explicitly specified

\maketitle      

\clearpage

%\linenumbers

\section{Abstract}
 Antiferroelectrics are antipolar materials which possess an electric field-induced phase transition to a polar, ferroelectric phase and offer significant potential for sensing/actuation and energy-storage applications. Known antiferroelectrics are relatively scarce and mainly based on a limited set of perovskite materials and their alloys (\textit{e.g.}, \ce{PbZrO3}, \ce{AgNbO3}, and \ce{NaNbO3}). Here, a new family of lead-free, weberite-type antiferroelectrics, identified through a large-scale, first-principles computational search is introduced. The screening methodology, which connects lattice dynamics to antipolar distortions, predicted that \ce{La3NbO7} could exhibit antiferroelectricity. We confirm the prediction through the synthesis and characterization of epitaxial \ce{La3NbO7} thin films, which display the signature double hysteresis loops of an antiferroelectric material as well as clear evidence of an antipolar ground state structure from transmission electron microscopy. The antiferroelectricity in \ce{La3NbO7} is simpler than most known antiferroelectrics and can be explained by a Kittel-type mechanism involving the movement of niobium atoms in an oxygen octahedron through a single phonon mode which results in a smaller change in the volume during the field-induced phase transition. Additionally, it is found that \ce{La3NbO7} combines a high threshold field with a high breakdown field ($\approx$ 6 MV/cm) $\textendash$ which opens up opportunities for energy-storage applications. This new weberite-type family of materials offers many opportunities to tune electrical and temperature response especially through substitutions on the rare-earth site. Ultimately, this work demonstrates a successful data-driven theory-to-experiment discovery of an entirely new family of antiferroelectrics and provides a blueprint for the future design of ferroic materials.

\section{Introduction}
Antiferroelectrics (AFEs) exhibit an antipolar ground state ($i.e.$, with electric dipoles adopting an antiparallel arrangement) that can be switched to a polar, parallel arrangement through application of an electric field \cite{Rabe2013, Catalan2025}. The field-induced transition from the antipolar to a polar phase gives rise to a characteristic double hysteresis loop in the polarization–electric field response. The switching of an AFE by the application of an electric field is comparable to the switching of an antiferromagnetic state to a ferromagnetic state but with electric dipoles instead of magnetic moments. The field-induced phase transition and the double-hysteresis loops present in AFEs are central to various applications ranging from energy-storage capacitors \cite{Luo2020}, electrocaloric solid-state cooling \cite{Geng2015}, actuators or transducers \cite{Pan2024}, or non-volatile random access memory \cite{Pevsic2016}. 

Antiferroelectricity is a (relatively) rare property found only in a handful of material families \cite{Randall2021}.  The prototypical AFEs are \ce{PbZrO3} (PZO) and lead-free \ce{NaNbO3} (NNO) and \ce{AgNbO3} (ANO), as well as their solid-solution derivatives \cite{Shirane1951, Yang2020}. To date, research on AFEs has focused predominantly on these materials, with relatively few systematic efforts devoted to identify new AFE compounds \cite{Bennett2013, banyas2025}. There has been, however, a resurgence of interest in AFE materials in recent years, as researchers have recognized the potential these materials can play in a range of applications and, as a result, the desire to search for new materials that provide the desired combinations of properties has grown \cite{Randall2021, Si2024}.

\begin{figure}
    \centering
    \includegraphics[width=0.9\linewidth]{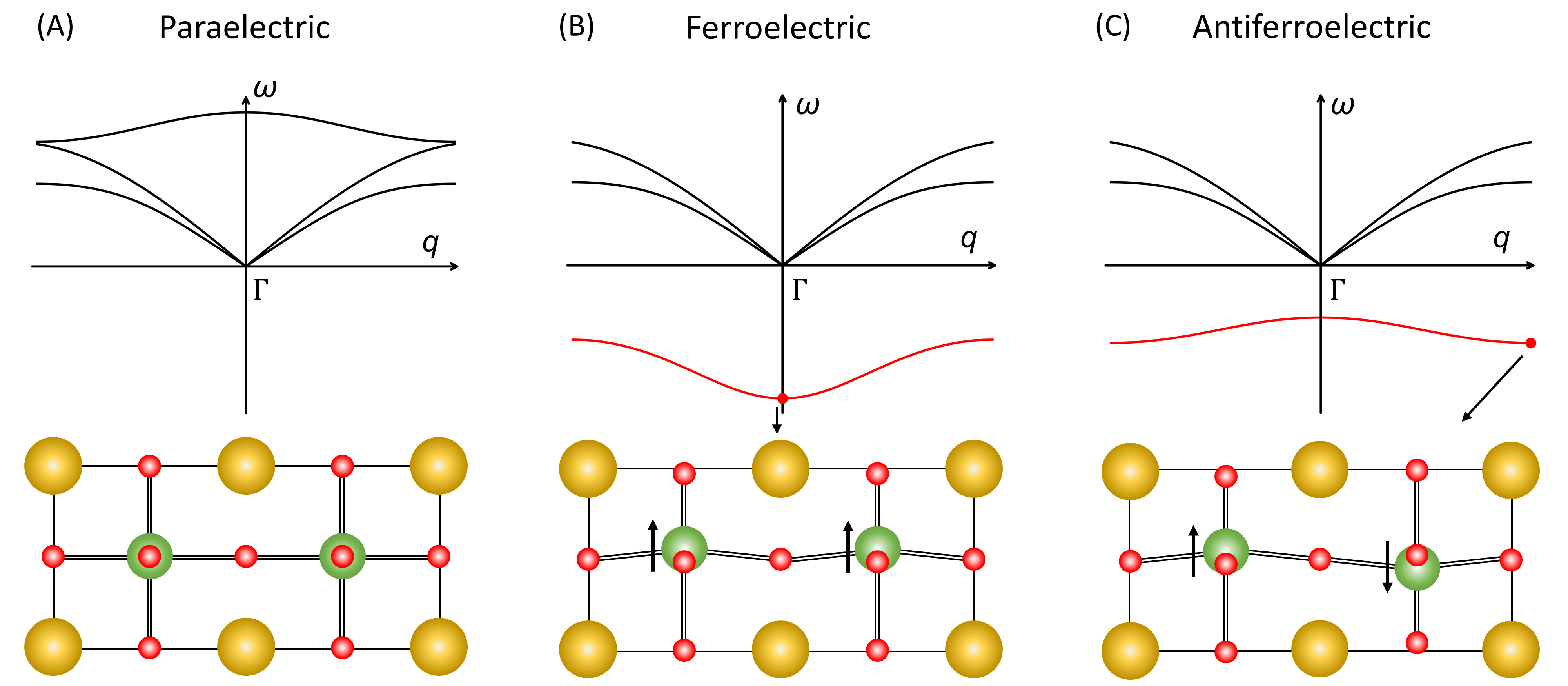}
    \caption{Schematic representation of the phonon band structures and the expected ground states for (A) a paraelectric material showing no instabilities and preserving a centrosymmetric structure, (B) a ferroelectric material with a dominant instability at $\Gamma$ and exhibiting a ferrodistortive pattern of distortion, and (C) an AFE material in which the dominant instability is found at a zone-boundary and thus, showing antipolar distortion. The arrows represents the atomic displacements which coincides with the local dipole moments. A perovskite-like crystal structure has been chosen for illustrative purposes. This represents an idealized case where the dominant instability leads to the energy ground-state. However, it is possible that the condensation of the mode with the largest imaginary frequency does not lead to the energy ground state.}
    \label{fig1_phonon}
\end{figure}

The soft-mode theory of ferroelectricity has been used to understand and even identify new ferroelectrics (FEs) using phonon band structures \cite{Garrity2018, Markov2021, He2024, Hirai2024}. Likewise,  antiferroelectrity can also be understood using phonons \cite{Rabe2013}. Most centrosymmetric materials exhibit a phonon band structure without any unstable modes indicating paraelectric (PE) behavior (\textbf{Fig. \ref{fig1_phonon}.A}).  The less common case of a centrosymmetric phase exhibiting an unstable mode in $\Gamma$, however, is an indicator of a stable polar distortion and of potential ferroelectricity (\textbf{Fig. \ref{fig1_phonon}.B}). Following this logic, the phonon band structure of a centrosymmetric material exhibiting AFE behavior should contain a dominant instability at a zone-boundary, corresponding to the antipolar phase while maintaining a competing unstable mode at $\Gamma$, corresponding to the polar phase (\textbf{Fig. \ref{fig1_phonon}.C}). In this case, the antipolar phase will be the ground state with a competing polar phase that could be favored by applying an electric field. It is noted that this is the simplest case possible for an AFE (often called Kittel-like AFEs) \cite{Kittel1951}. More complex behavior ($e.g.$, involving the condensation of several phonon modes to form the (anti-)polar phase) are commonly observed in FEs and AFEs \cite{Iniguez2014, Amisi2023, Rabe2013, Catalan2025}.

Following this idea from soft-mode theory, we use here a first-principles database of phonons to identify potential AFEs among approximately 5,000 materials. Specifically, we computationally identify a new family of AFEs in the weberite-type \ce{\textit{A}$_3$\textit{M}O$_7$} (\textit{A} = a rare-earth element, \textit{M }= Nb, Ta) structure. We confirm our computational prediction through thin-film synthesis and characterization of \ce{La3NbO7}. Our experiments confirm the antipolar nature of the AFE ground state using scanning transmission electron microscopy (STEM) and demonstrate electrical switching with a double-hysteresis. The new \ce{La3NbO7} phase shows a transition between the antipolar (AFE) and the polar (FE) phases that is Kittel-like and simpler than PZO, NNO, or ANO. \ce{La3NbO7} exhibits unique properties such as high threshold and breakdown fields as well as small volume changes during switching. Beyond \ce{La3NbO7}, an entire family of weberite-type AFEs is uncovered and it is shown that there is potential for tuning the electrical response through alloying on the \textit{A} and \textit{M} sites.

\section{Results}
\subsection{Searching for AFE materials in a phonon database}

Our search for new AFEs starts with a database of DFT phonon band structures for 4,957 centrosymmetric oxides \cite{Phonopy2015, Togo2018, Phonopy2023} present in the \textit{Materials Project} \cite{Jain2013}. Using this database, we found 696 materials which exhibit imaginary modes, a signature of structural instabilities. For each of these materials with instabilities, that will be referred to as \textit{parents}, we obtained a series of structural distortions by condensing the unstable modes which we subsequently relaxed using density functional theory (DFT) calculations, obtaining what will be referred to as \textit{children} structures (Methods). As outlined in \textbf{Fig. \ref{fig1_phonon}}, we expect a potential AFE to emerge if a non-polar ground state emerges from a dominant zone-boundary instability, with a polar phase originating from a $\Gamma$-centered mode marginally higher in frequency. Indeed, AFEs need to have a ground state that is non-polar formed by alternating antiparallel dipoles and a polar distortion close enough in energy so an electric field can be applied to switch the AFE ground state to a FE state. We note that our screening relies on phonon band structure to identify potential candidates and suggest modes to relax along but it is the energetics of the competing distortions (polar, anti-polar AFE, non-polar PE) after relaxation that is ultimately used to identify new AFEs.

For the 27 materials offering an AFE ground state, we gather two important proxies that we will use to identify their potential as a switchable AFE material: $\Delta E_{parent}$ and the threshold field $F_{th}$. $\Delta E_{parent}$ is the difference in energy between the original, centrosymmetric, PE, parent phase and the AFE phase. This energy $\Delta E_{parent}$ is related to both the barrier to switch the material and to the transition temperature (Curie temperature, $T_C$) at which the AFE phase transforms to the parent PE phase.\cite{Wojdel2014, Zhang2017, Ricci2024} Ideally, $\Delta E_{parent}$ should be neither too low, to avoid a too small $T_C$, nor too high as it can translate to a high barrier to switch the AFE.  While there is no hard cut-off as switching is a complex process, Ricci \textit{et al}. have shown that switchable FEs have usually a $\Delta E_{parent}$ lower than 100 meV/atom \cite{Ricci2024}.

The other parameter to look at is the threshold field ($i.e.$, the electric field at which the AFE phase becomes FE):  $F_{th}=\frac{\Delta E_{AFE-FE}}{P \cdot \Omega_0}$ where $\Delta E_{AFE-FE}$ is the energy difference between the AFE and FE phase, $P$ is the polarization of the ferroelectric phase, and $\Omega_0$ is the unit-cell volume. If $F_{th}$ is too small, the double-hysteresis may not be clearly developed ($i.e.$, the material would behave almost as a FE) and if it is too large, the stabilization of the FE phase would require fields that are too high and could go beyond the material's electrical-breakdown voltage. We note that the mechanism by which AFEs switch under field or experience phase transitions with temperature can be complex, involving multi-scale effects. $\Delta E_{parent}$ and $E_{th}$ are used here as proxies for experimental $T_C$, threshold field, and/or switching barrier. More computational or experimental work is needed to refine the exact AFE properties of these potential candidates.

\begin{figure}
    \centering
    \includegraphics[width=0.75\linewidth]{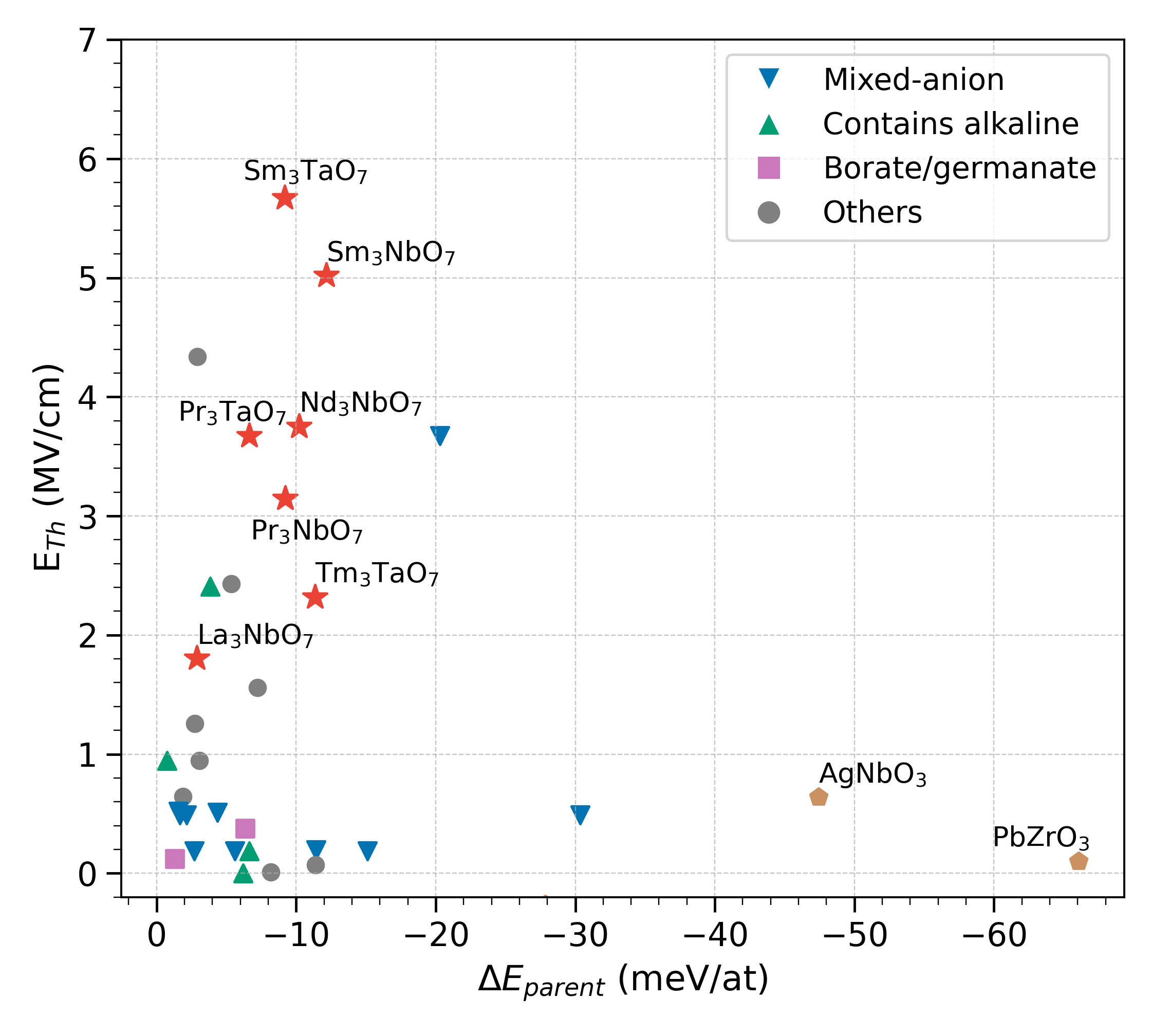}
    \caption{Energy difference between the parent and AFE phase $\Delta E_{parent}$ (in meV/atom) versus threshold field $E_{th}$ (in MV/cm) for all AFE candidates identified through our high-throughput screening as well as for two prototypical AFEs (PZO and ANO). All calculations have been performed using PBEsol. The color and markers indicate different chemistry and classes of materials.}
    \label{fig2_AFE_HT}
\end{figure}

In \textbf{Fig. \ref{fig2_AFE_HT}}, we plot the computed values for $\Delta E_{parent}$ and $F_{th}$ for our AFE candidates and for two prototypical AFEs: PZO and ANO. A list of a few known AFEs as well as all the candidates found during our search with their computed properties is available in Supplementary Table \ref{SM_tab:known_AFEs} and \ref{SM_tab:AFE_candidates}. We grouped materials in chemical classes taking into account synthesizability: alkaline-containing materials (green triangles) and mixed-anion (blue triangles) are usually not the most amenable to traditional thin-film-growth techniques. We also found a series of borate/germanate materials ($\ce{Ba3Ti3(BO3)2O6}$, $\ce{K3Nb3(BO3)2O6}$, $\ce{K3Ge3(BO3)2O7}$, indicated by purple squares, which combine low $F_{th}$ with high $\Delta E_{parent}$ and in which (anti-)ferroelectricity has been experimentally reported \cite{Becker1996, Shan2020}. 

While PZO and ANO sit at high $\Delta E_{parent}$ and low $F_{th}$, we find many materials exhibiting opposite behavior with higher $F_{th}$ but lower $\Delta E_{parent}$. Higher threshold field could be beneficial for energy-storage applications provided the breakdown field is high as well. We note that the higher $\Delta E_{parent}$ of PZO and ANO could be related to their more complex switching mechanism. Among our candidates, we find a very interesting family of materials of general formula \ce{\textit{A}$_3$\textit{M}O$_7$} forming in the weberite-type structure and marked by red stars (\textbf{Fig. \ref{fig2_AFE_HT}})\cite{Cai2009}. We provide their phonon band structure (Supplementary \textbf{Fig.} \ref{SM_fig:A3MO7_phonon_1} and \ref{SM_fig:A3MO7_phonon_2}). The few members identified in the database cover a wide range of $\Delta E_{parent}$, from -2.87 meV/atom in \ce{La3NbO7} to -12.18 meV/atom in \ce{Sm3NbO7} and $F_{th}$, which spans from 1.80 MV/cm in \ce{La3NbO7} to 5.67 MV/cm in \ce{Sm3TaO7}. While direct translation of these values from DFT to experimental measurement is difficult, \textbf{Fig. \ref{fig2_AFE_HT}} definitely indicates that the weberite-structure offers a new type of antiferroelectricity with a potentially simpler switching mechanism, lower barrier for switching ($\Delta E_{parent}$), and higher threshold field $F_{th}$. Importantly, the presence of several members of the same structural family shows the potential for alloying between rare-earth on the \textit{A} site and/or niobium and tantalum on the \textit{B} site as a way to tune the energy landscape. This alloying strategy is commonly used to optimize the performance of perovskite ferroic systems (\textit{e.g.}, PbZr$_{1-x}$Ti$_x$O$_3$, (1-x)PbMg$_{1/3}$Nb$_{2/3}$O$_3$-(x)PbTiO$_3$, or \ce{NaNbO3}-\ce{AgNbO3}) and is a clear advantage for the weberite-structure. \cite{Haertling1999, Yang2020} 

One of the applications for AFEs is in capacitive-energy storage and, assuming an ideal, adiabatic switching, the stored energy follows the product of the threshold field and the polarization \cite{Gaur2024}. While the weberite-structures show lower polarizations than perovskites such as PZO and ANO, they can make up for this in energy storage capacity through their higher threshold field (provided their breakdown field is high) (Supplementary \textbf{Fig.} \ref{SM_fig:P_vs_Eth}). The weberite-like structures open a new path towards high threshold field AFEs for energy storage, complimentary to the high polarization-low threshold field offered by PZO and ANO.

\subsection{Antiferroelectricity and phase competitions in \ce{La3NbO7}}
\begin{figure}
    \centering
    \includegraphics[width=0.9\linewidth]{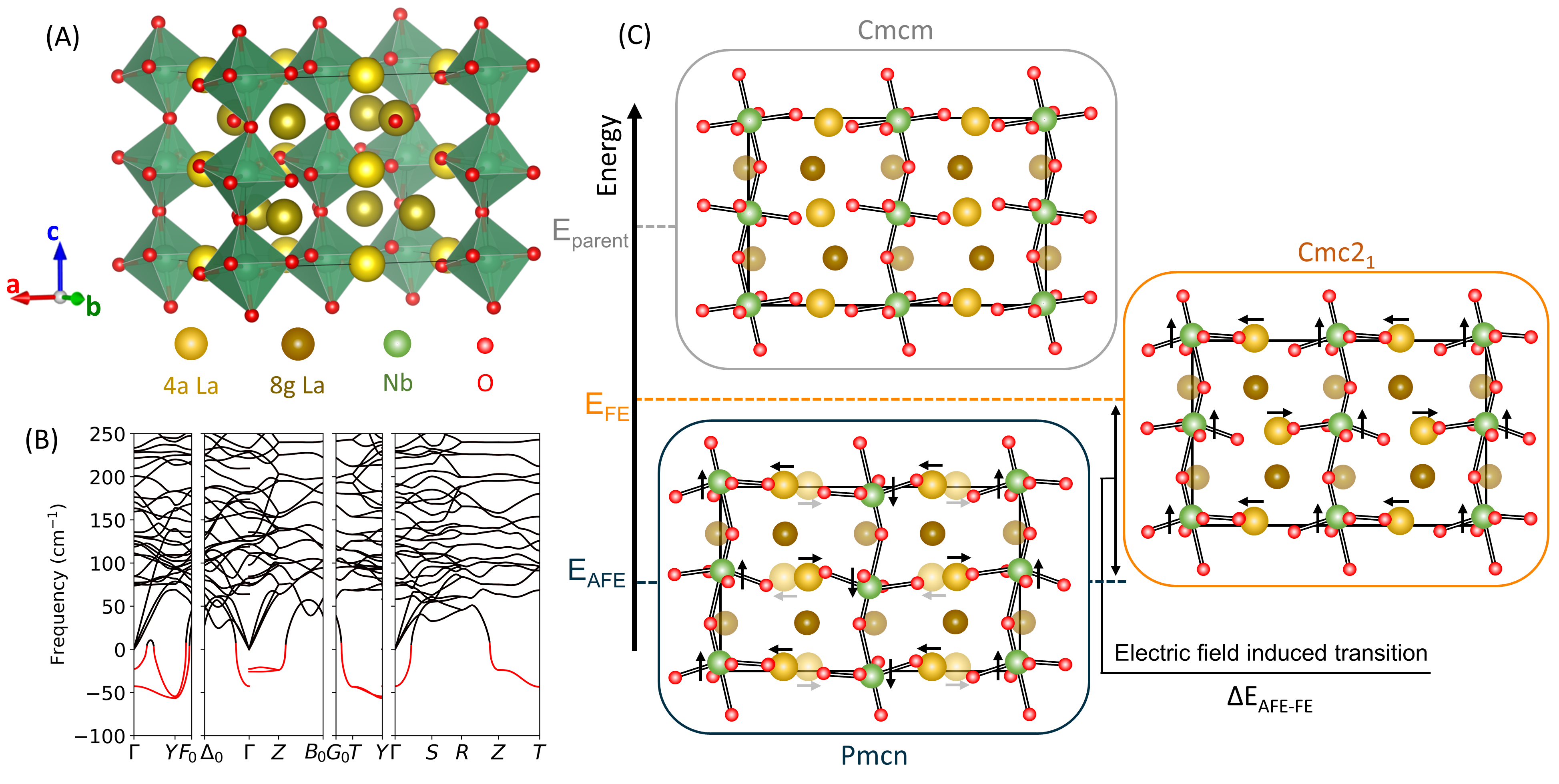}
    \caption{(A) Parent structure of \ce{La3NbO7}, which is paraelectric and belongs to the \textit{Cmcm} space group. The \ce{NbO6} octahedra are shown in dark green while we distinguish the layers of 4a (light gold) and 8g (dark gold) lanthanum atoms. (B) Phonon band structure of \ce{La3NbO7}. The main instabilities are found at $\Gamma$ and $Y$, with the latter being slightly more negative (imaginary). (C) Structures of the AFE (\textit{Pnma}), FE (\textit{Cmc}2$_1$) and PE (\textit{Cmcm}) phases of \ce{La3NbO7} and their respective energies. The displacements of the atoms in the AFE and FE phases with respect to the PE phase are shown by dark arrows.}
    \label{fig3:DFT_structures}
\end{figure}

While we identified a series of weberite-type materials, we focus first on \ce{La3NbO7} which shows the lowest $E_{th}$ and $\Delta E_{parent}$. These low values suggest the possibility for lower-field AFE switching, hopefully, below the breakdown field. The centrosymmetric phase of \ce{La3NbO7} forms in the orthorhombic weberite-type structure with space group \textit{Cmcm} (\textbf{Fig. \ref{fig3:DFT_structures}A)} \cite{Allpress1979, Rossell1979}. The main structural feature of this PE-parent structure is the presence of NbO$_6$ octahedra (in dark green,  \textbf{Fig. \ref{fig3:DFT_structures}A}) chains along the [001] and presenting a significant tilt around the [100]. The lanthanum occupies two distinct sets of Wyckoff position, 4a (light gold) and 8g (dark gold).  We plot the phonon band structure of \ce{La3NbO7}  (\textbf{Fig. \ref{fig3:DFT_structures}B}) and observe unstable modes dominated by the antipolar zone-boundary at $Y$ and a polar mode at $\Gamma$ slightly higher in frequency. These features in the phonon band structure are compatible with a Kittel-like AFE (\textbf{Fig. \ref{fig1_phonon}}) explaining why the material was selected by our screening. The unstable modes at $Y$ lead to an AFE ground state in a \textit{Pnma} space group while the mode at $\Gamma$ leads to a \textit{Cmc}2$_1$ FE structure (\textbf{Fig. \ref{fig3:DFT_structures}C}, Supplementary \textbf{Fig.} \ref{SM_fig:In-plane_structures} and Supplementary Note \ref{sm_note:ground_state}). In both structures, the distortion involves the off-centering displacement of the niobium atoms within their octahedra along the [001]. In the \textit{Pnma}, AFE phase, the displacements of the niobium in neighboring chains is antipolar while they are in-phase in the FE, \textit{Cmc}2$_1$ phase, thus leading to a calculated polarization of  $\approx$ 8.5 $\mu C/cm^2$ along the [001]. The transition from the AFE to the FE phase is, thus, mostly due to the shift of the Nb atoms and does not involve any octahedral rotations as in PZO, NNO, and ANO \cite{Iniguez2014, Luo2023}. We also note that there is an antipolar motion of neighboring (001) of the 4a lanthanum atoms essentially along the [010]. The displacements of the atoms within a given plane are ferrodistortive in the FE phase and antiferrodistortive in the AFE phase. The 8g lanthanum atoms do not change position between the different phases. Both the parent and AFE phases have been observed experimentally \cite{Cai2011} (as expected as our screening focuses on known phases), but we are (to the best of our knowledge) the first to report the existence of the FE, \textit{Cmc}2$_1$ phase and to identify its AFE potential.

\subsection{Experimental demonstration of antiferroelectrity in \ce{La3NbO7}}
\begin{figure}
    \centering
    \includegraphics[width=0.86\linewidth]{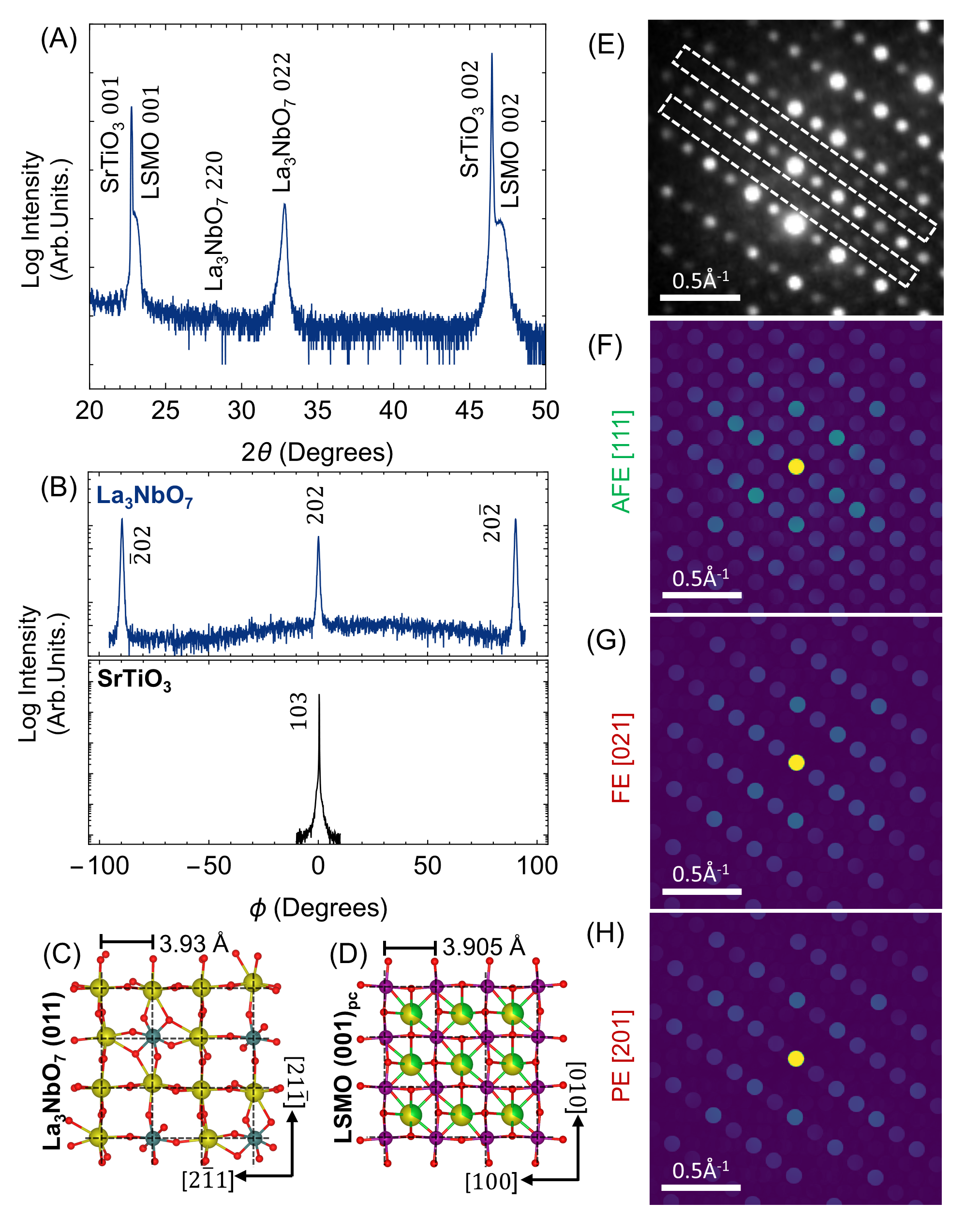}
    \caption{X-ray- and transmission electron microscopy (TEM)-based diffraction measurements from \ce{La3NbO7} films. \textbf{a.} X-ray $\theta$-2$\theta$ linescans show films are predominately (011)-oriented, while \textbf{b.} $\phi$-scans establish the epitaxy with the underlying (pseudo)subic LSMO electrode and \ce{SrTiO3} substrate, where \textbf{c.} an approximately square lattice is formed on the \ce{La3NbO7} (011) interface whose lattice parameter is relatively close to \textbf{d.} that of LSMO strained to \ce{SrTiO3}. \textbf{e.} 4D-STEM nanobeam diffraction demonstrates obvious 1/2-order spots which agree with \textbf{f.} the simulated electron diffraction pattern of the AFE [111] zone axis and are distinct from the alternative \textbf{g.} FE [021] and \textbf{h.} PE [201] zone axes.}
    \label{fig4:La3NbO7_Structure}
\end{figure}

Following the computational prediction of potential AFE behavior, \ce{La3NbO7} was synthesized as a thin film and characterized structurally and electrically. Thin films of \ce{La3NbO7} were grown via pulsed-laser deposition (Methods) on \ce{La$_{0.7}$Sr$_{0.3}$MnO$_3$} (LSMO) electrodes which were coherently strained to the underlying \ce{SrTiO3} (001) substrates, and their structure was verified via X-ray diffraction (Methods). $\theta$-2$\theta$ linescans show predominately (011)-oriented films with faintly detectable 110-type diffraction peaks (\textbf{Fig. \ref{fig4:La3NbO7_Structure}A}). Based on azimuthal ($\phi$) scans about the 202-diffraction condition for (011)-oriented films (\textbf{Fig. \ref{fig4:La3NbO7_Structure}B}), the \ce{La3NbO7} 202-diffraction condition is found at the same rotation angle as the substrate 103-diffraction condition ($i.e.$, the \ce{La3NbO7} [21$\overline{1}$] is parallel to the \ce{SrTiO3} [100]). This yields a relatively straightforward epitaxy in which the pseudo-square lattice of the \ce{La3NbO7} cations (\textbf{Fig. \ref{fig4:La3NbO7_Structure}C}) aligns with the pseudo-square lattice of the electrode cations (\textbf{Fig. \ref{fig4:La3NbO7_Structure}D}). The average lattice parameter of this lattice for \ce{La3NbO7} is $\approx$ 3.93 Å, which is relatively close to that of \ce{SrTiO3} (\textit{a} = 3.905 Å). 

Given the similarities in structure and lattice parameters between AFE, FE, and PE phases of \ce{La3NbO7}, the predicted AFE structure cannot be uniquely verified with X-ray diffraction alone from an epitaxial thin film. To further differentiate these structures, four-dimensional STEM (4D-STEM) nanobeam diffraction (Methods) was conducted on the \ce{La3NbO7} films (\textbf{Fig. \ref{fig4:La3NbO7_Structure}E} and Supplementary Note \ref{sm_note:tem}). Depending upon the film structure, the zone axis measured could either be the AFE [111] (\textbf{Fig. \ref{fig4:La3NbO7_Structure}F}), FE [021] (\textbf{Fig. \ref{fig4:La3NbO7_Structure}G}), or PE [201] (\textbf{Fig. \ref{fig4:La3NbO7_Structure}H}). While these FE and PE zone axes are indistinguishable, the AFE [111] zone axis should reveal 1/2-order diffraction spots (indicative of the antipolar ground state) in its pattern which are not present in the other structures (\textbf{Fig. \ref{fig4:La3NbO7_Structure}G,H}). These superlattice spots can clearly be seen in the 4D-STEM nanobeam diffraction data and the simulated diffraction pattern (highlighted in the dashed boxes, \textbf{Fig. \ref{fig4:La3NbO7_Structure}\textbf{E},\textbf{F}}), providing strong evidence that the observed structure is AFE. Additionally, the AFE and FE structures have small lanthanum distortions which are not present in the PE phase and can be observed in real space along the AFE [010]/FE [001] (corresponding to the PE [001]). Atomic-resolution high-angle annular dark-field (HAADF) STEM images taken from the \ce{La3NbO7} films (\textbf{Fig. \ref{fig:tem}}) do show faint indications of such distortions, further supporting the presence of the AFE phase. Thus, taken together, the indications of the appropriate distortions from the real-space images and the presence of the 1/2-order peaks in the diffraction measurements strongly supports the formation of the expected antipolar, AFE, phase.

\begin{figure}
    \centering
    \includegraphics[width=1.0\linewidth]{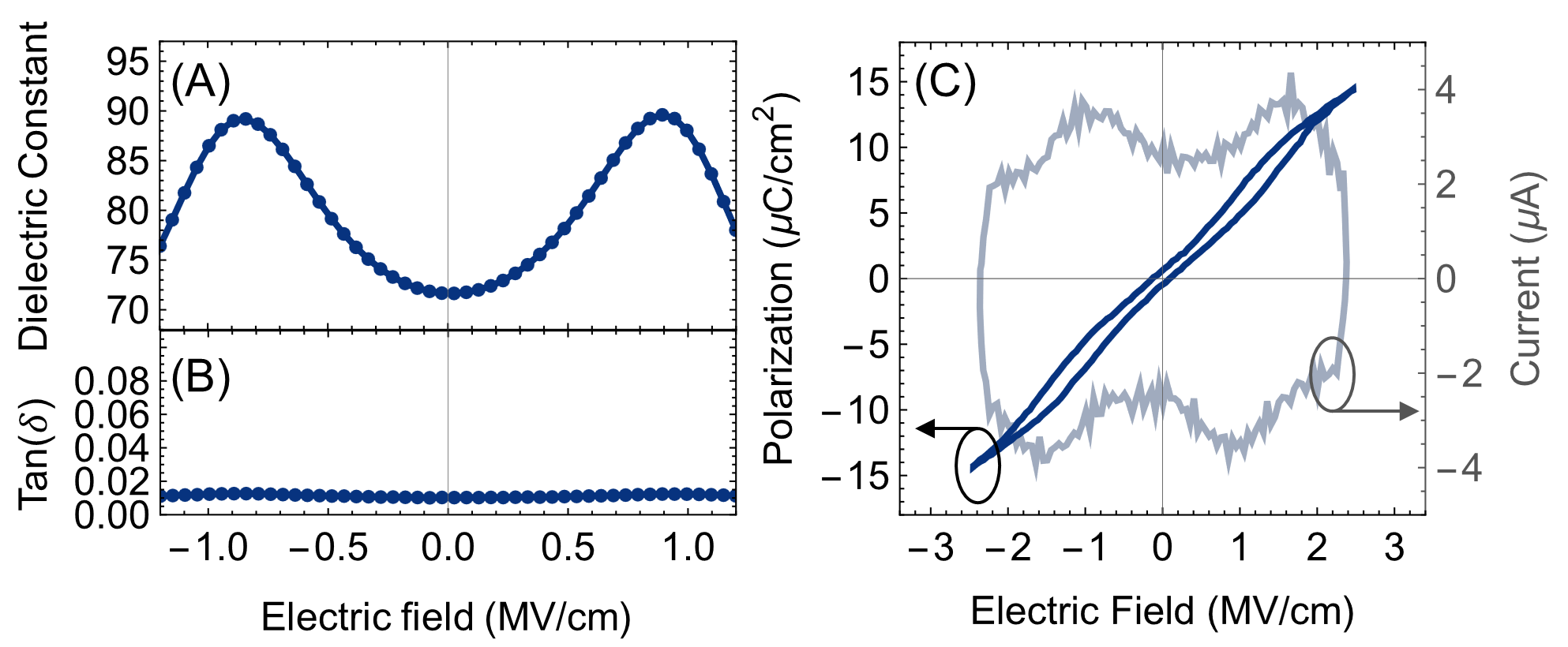}
    \caption{Capacitor-device-based \textbf{a.} dielectric constant and \textbf{b.} loss tangent versus DC electric field and \textbf{c.} polarization (left axis) and current (right axis) versus electric field measured at 10 kHz. All measurements demonstrate strong indications of the expected AFE nature, with double peaks in capacitance measurements, double hysteresis in polarization, and four peaks in current response.}
    \label{fig5:La3NbO7_Properties}
\end{figure}

Following analysis of the crystal structure, capacitor-based device structures were fabricated from symmetric LSMO/\ce{La3NbO7}/LSMO/\ce{SrTiO3} (001), thin-film heterostructures to evaluate their functional properties (Methods). At room temperature, all films were found to be AFE, based on capacitance versus voltage measurements which show two distinct peaks, corresponding to the positive and negative critical fields for the AFE-to-FE transition (\textbf{Fig. \ref{fig5:La3NbO7_Properties}A}), while dielectric loss (\textbf{Fig. \ref{fig5:La3NbO7_Properties}B}) remains low, suggesting the peaks are not the result of leakage at higher electric fields. Polarization versus electric-field measurements taken at room temperature (\textbf{Fig. \ref{fig:loops}A-D} and Supplementary Note \ref{sm_note:hysteresis}) show very slight curvature which, by itself, is not a strong indication of antiferroelectricity, but measurements taken at lower temperatures (12 K) show more classical AFE double-hysteresis (\textbf{Fig. \ref{fig5:La3NbO7_Properties}C}). Additionally, current loops taken at room temperature (\textbf{Fig. \ref{fig:loops} A-D}) and 12 K (\textbf{Fig. \ref{fig:loops}E-H} and \textbf{Fig. \ref{fig5:La3NbO7_Properties}C}) show four distinct peaks, further verifying the AFE nature. The lack of sharp hysteresis at room temperature may arise due to a relatively shallow energy barrier between the AFE and FE phases, low polarization, and/or proximity to the phase transition temperature. The peaks in Fig. \textbf{\ref{fig5:La3NbO7_Properties}C} are rather broad but indicate a high threshold field between 1 and 2 MV/cm in agreement with the computational prediction of 1.8 MV/cm. This threshold field is remarkably higher than in ANO and PZO which have shown experimental threshold fields from 0.15 to 0.3 MV/cm \cite{Luo2020, Pan2024} (Supplementary Table \ref{SM_tab:known_AFEs}).

To further investigate the phase transition in this system, dielectric response was measured as a function of temperature (\textbf{Fig. \ref{fig:phase-transition}A-B} and Supplementary Note \ref{sm_note:exp_phase_transition}) and it shows a relatively broad peak at 430-450 K. This transition is further verified through capacitance versus voltage measurements (\textbf{Fig. \ref{fig:phase-transition}C-D}), which show the two antiferroelectric peaks weakening and eventually joining together near the transition temperature. To further characterize the nature of this phase transition, temperature-dependent measurements were fit to modified Curie-Weiss behavior: \begin{equation}
\label{eq:Curie-Weiss}
    \frac{1}{\varepsilon} - \frac{1}{\varepsilon_{max}} = (T - T_{max})^{\frac{\gamma}{C}}
\end{equation} where $\varepsilon$ is the dielectric constant, $\varepsilon_{max}$ is the maximum dielectric constant (occurring at the phase transition),  $T$ is the temperature, $T_{max}$ is the temperature at the phase transition, $C$ is the Curie constant (a fitting parameter), and $\gamma$ is a measure of the diffusivity of the transition \cite{Du2009}. For ferroelectric materials, ideal behavior should produce a $\gamma \approx$ 1 while an ideal relaxor would produce a $\gamma \approx$ 2. For \ce{La3NbO7}, $\gamma \approx$~1.27 (\textbf{Fig. \ref{fig:phase-transition} E}), indicating the phase transition is partially diffuse but remains relatively ideal and ferroelectric-like. 

In addition to the expected antiferroelectric properties, the \ce{La3NbO7} films also show exceptional breakdown performance. To characterize the high-electric-field performance, Weibull distribution measurements were collected from 15 devices (\textbf{Fig. \ref{fig:breakdown}} and Supplementary Note \ref{sm_note:breakdown}) by applying a sequence of monopolar polarization versus electric-field measurements, incrementing the voltage for each loop, and recording the breakdown field (defined by an open- or short-circuit in the measurement and a visible cracking in the device) (Methods). From these measurements, a characteristic breakdown field of $\approx$ 6 MV/cm was extracted. This is significantly higher than many optimized AFEs. PZO shows breakdown fields from 0.1-2.5 MV/cm and lower values have been reported for NNO and ANO \cite{Chauhan2025, Nguyen2021, Fang2021, Chen2020, Yang2025, Zhao2023, Ma2023}. While intrinsic breakdown fields have been often linked to the band gap, PZO and \ce{La3NbO7} have very similar band gaps with computed values $\approx$ 4.15 eV using the HSE functional (Methods), in agreement with PZO experimental band gap of 3.6-3.7 eV \cite{Tang2000, Moret2002}. In fact, empirical relations between band gap and breakdown field would indicate that a 4 eV band gap materials should have intrinsic breakdown fields between 4 and 10 MV/cm \cite{Higashiwaki2012, Wang2006}. This hints to the importance of extrinsic factors in the large breakdown field difference between PZO and \ce{La3NbO7}.

\section{Antiferroelectricity in the weberite-like family \ce{\textit{A}$_3$\textit{M}O$_7$}}

\begin{figure}
    \centering
    \includegraphics[width=0.9\linewidth]{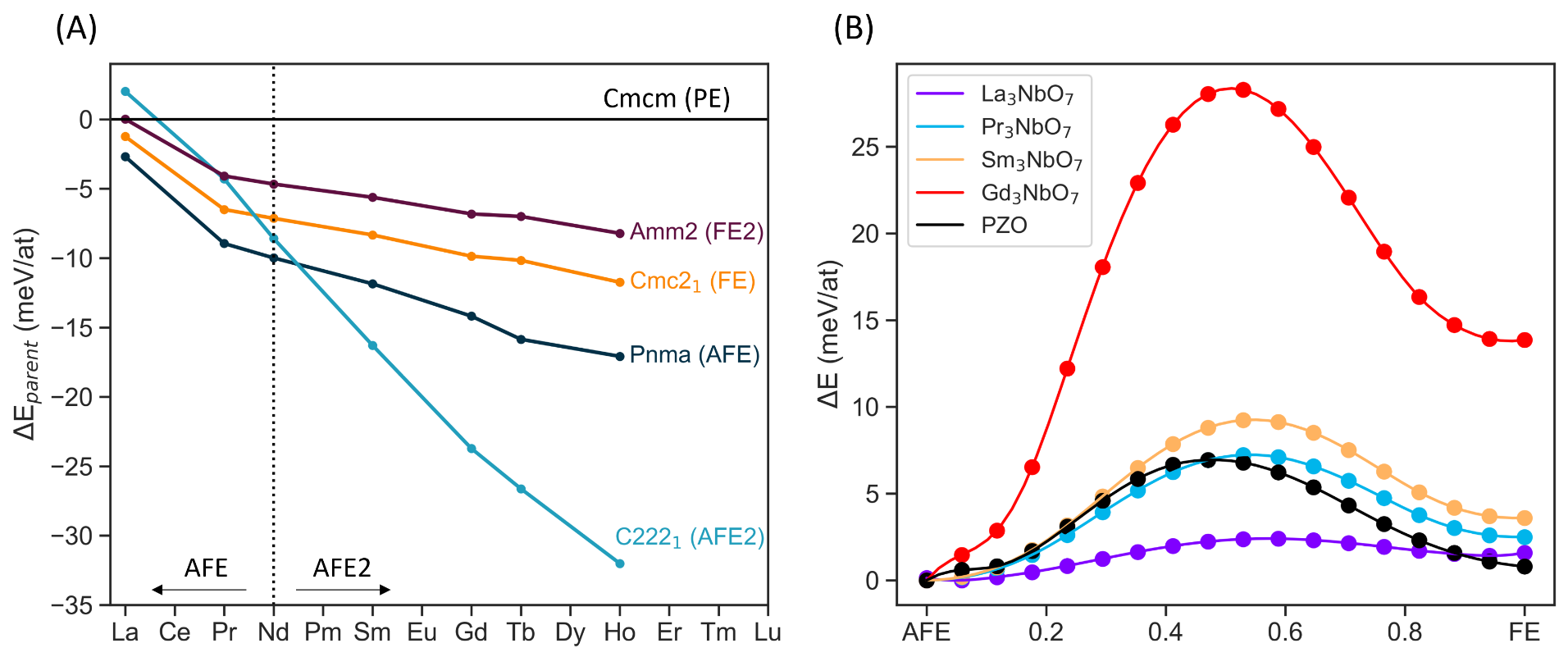}
    \caption{(A) Energy landscape of \ce{\textit{A}$_3$NbO$_7$}, given with respect to the parent phase, across the lanthanide series. We indicated the AFE (\textit{Pnma}), FE (\textit{Cmc}2$_1$), AFE2 (\textit{C}222$_1$) and FE2 (\textit{Amm}2) phases with solid lines. The filled region represents the experimentally observed phases at room temperature as found in \cite{Doi2009, Wakeshima2010}. (B) Energy barriers as calculated by the NEB method for the AFE to FE transition for a few representative of the lanthanide series, as well as for PZO.}
    \label{fig4:DFT_energies}
\end{figure}

We have demonstrated using theory and experiments that \ce{La3NbO7} is a new AFE. Our initial screening indicated that other compositions in the same weberite-like structure could be AFEs as well (\textbf{Fig. \ref{fig2_AFE_HT}}). Beyond these other phases themselves, this has important consequences for the ability to tune key functional properties, including the transition temperature $T_C$ and the threshold field in \ce{La3NbO7}, via alloying. We study now in more detail how the energetics of phase stability of the \ce{\textit{A}$_3$\textit{M}O$_7$} system is influenced by rare-earth substitution. We focus on the niobium series but the conclusions for tantalum are similar (Supplementary Table \ref{SM_tab:energy_landscape_Ta}). In \textbf{Fig. \ref{fig4:DFT_energies}}, we plot the energy of the \textit{Pnma} (AFE) and \textit{Cmc}2$_1$ (FE) structure with respect to the parent \textit{Cmcm} (PE) used as a reference. We also add a \textit{C}222$_1$ (AFE$_2$) antipolar structure which was missing from our initial search and was instead found from literature \cite{Allpress1979, Doi2009, Chesnaud2015, Chen2018}. In this structure, the NbO$_6$ octahedra chains are tilted about the [010] rather than the [100] as in all the aforementioned phases and cannot be described by the condensation of a single phonon mode (Supplementary \textbf{Fig.} \ref{SM_fig:difference_between_C2221_Pnma}) \cite{Gussev2023}. Additionally, for rare-earth elements smaller than lanthanum, we find a higher energy additional phase with space group \textit{Amm}2 in which the neighboring (001) of 4a lanthanum atoms move in-phase, leading to a large in-plane polarization of $\approx$ 30  $\mu C/cm^2$ along the [010] (Supplementary \textbf{Fig.} \ref{SM_fig:In-plane_structures}).

For the largest rare-earths (lanthanum to neodymium), the calculated ground state is the \textit{Pnma}, AFE phase (dark blue line,  \textbf{Fig. \ref{fig4:DFT_energies}A}), and this phase is stabilized with respect to the parent phase as one moves along the series (towards smaller rare-earths). The energy difference between the AFE and FE phases (orange line, \textbf{Fig. \ref{fig4:DFT_energies}A}), also increases for smaller rare-earths. This suggests that alloying between rare-earths could tune the energy landscape. For smaller rare-earths starting at neodymium, the \textit{C}222$_1$ structure becomes the ground state (light blue line,  \textbf{Fig. \ref{fig4:DFT_energies}A}). Even though it also displays antipolar displacements, it is unlikely that switching the \textit{C}222$_1$ (AFE$_2$) to the FE phase will be as easy as for the \textit{Pnma} (AFE$_1$) phase due to the high energy cost required to reorient the NbO$_6$ octahedral chains. In view of these results, the largest rare-earths (lanthanum to neodymium) in the \ce{\textit{A}$_3$NbO$_7$} series are the most likely to present switchable AFE materials. 

The crystallography of the \ce{\textit{A}$_3$NbO$_7$} family has been studied in the past, enabling us to compare our calculations with experimental results \cite{Allpress1979, Rossell1979, Kahn1995, Doi2009, Wakeshima2010, Cai2011, Chesnaud2015}. A summary of the different reported structures for \textit{A} = La, Nd, Eu, Gd, Ho, and Lu is provided (Supplementary Table \ref{SM_tab:A3NbO7_experimental_structure_summary}). For the early rare-earths (lanthanum and neodymium), the phases are both reported to be in the \textit{Cmcm} (PE) and \textit{Pnma} (AFE) space groups indicating the challenge in detecting the small AFE distortion. On the other hand, intermediate rare-earths (promethium to terbium) are reported as \textit{C}222$_1$ in agreement with DFT. Interestingly, previous work on \ce{Gd3NbO7} suggested that this material could be AFE but electric-field-induced switching could not be demonstrated \cite{Cai2010}. This is in agreement with the large calculated energy difference between the \textit{C}222$_1$, AFE and FE phases. Further computations of the switching barrier using the Nudged Elastic Band (NEB) approach confirm this finding and indicate small barriers (from 3 to 9 meV/atom) for the AFE (\textit{Pnma}) to FE switching for the early rare-earths (lanthanum, sumarium, and praseodymium) and a significantly larger barrier ($\approx$ 25 meV/atom) for gadolinium (switching the AFE \textit{C}222$_1$ phase to FE). For comparison, the computed barriers for the \textit{Pbam} to \textit{R}3\textit{c} transition in PZO with the same functionals are $\approx$ 7 meV/atom. Finally, we note that rare-earths smaller than terbium are expected to crystallize in the defective-fluorite \textit{Fm}$\bar{3}$\textit{m} \cite{Doi2009}.

While the \textit{Pnma} AFE structure is the ground state in the small rare-earths, increasing temperature can lead to a transition to the parent, PE phase. For \ce{La3NbO7}, we observed a phase transition from AFE to PE using \textit{ab initio} molecular dynamics (AI-MD) simulations by monitoring the AFE order parameter (Supplementary \textbf{Fig. \ref{SM_fig:AIMD}} and Supplementary Note \ref{sm_note:transition_dynamics}). We note that this is a qualitative observation and that an accurate computation of $T_C$ would require larger cell size and longer simulations \cite{Zhang2023}. We expect this transition temperature to increase as the rare-earth gets smaller and $\Delta E_{parent}$ increases. This is consistent with small anomalies observed in heat capacity measurements at 340-370 K and 430-470 K for \ce{La3NbO7} and \ce{Nd3NbO7}, respectively, as these anomalies were assigned to a transition from the AFE, \textit{Pnma} to the PE, \textit{Cmcm} phase \cite{Cai2011}, (in agreement with our own temperature-dependent dielectric measurements, Supplementary \textbf{Fig.} \ref{fig:phase-transition} and Note \ref{sm_note:exp_phase_transition}). 

\section{Discussion}
The discovery of \ce{La3NbO7} as a switchable, room-temperature antiferroelectric (AFE) expands the antiferroelectric materials landscape beyond the small set of prototypical systems (i.e., PZO, NNO, and ANO) that have dominated prior work. In this context, \ce{La3NbO7} is particularly distinctive, as it represents a Kittel-type AFE in which the transition from the PE to the AFE phase is driven by a single antipolar mode \cite{Kittel1951, Rabe2013, Catalan2025}. Such a Kittel-type transition was recently observed in the antipolar francisite \ce{Cu3Bi(SeO3)2O2Cl} but an electric field-induced transition to a polar phase could not be obtained \cite{Milesi2020}. In fact, as far as we know, \ce{La3NbO7} is the first experimentally demonstrated switchable Kittel-type AFE. 

The simplicity of the transition to the polar state, which is associated with the realignment of the niobium sublattices within their octahedral cages and the in-plane displacements of the lanthanum atoms, contrasts with the situation in PZO. In PZO, the AFE \textit{Pbam} structure is obtained from the PE phase by the condensation of several phonon modes. One mode ($\Sigma_2$) mostly associated with the antipolar displacement of the lead sublattice and another ($R_5^-$) involving the antiphase rotation of the \ce{ZrO6} octahedra \cite{Iniguez2014, Zhang2024}. Similar observations are made in the NNO and ANO perovskites \cite{Lin2024}. The detilting of the octahedra in PZO, NNO, and ANO induces a large volume change during the electric field-induced transition through the auxetic effect and has been used to obtain significant electromechanical response \cite{Fu2007, Pan2024, Lin2024}. On the contrary, in \ce{La3NbO7}, our calculations show that the volume changes very minimally upon transition with an expansion of +0.07\% along the polar axis and changes of -0.51\% and +0.16\% in the in-plane directions. This could be beneficial for cycling as large volume change has been linked to cracking and challenges in reversibility \cite{Zhang2021, Rabe2013}. In addition, \ce{La3NbO7} possesses a high threshold field (1 to 2 MV/cm) combined with a large breakdown field of $\approx$ 6 MV/cm and is alkaline- and lead-free. The significantly higher threshold and breakdown fields than current AFEs ($e.g.$, PZO, ANO or NNO) for an unoptimized film are of great interest and can be beneficial for applications such as energy storage.

There is much room for improvement in the results provided on unoptimized films. Reducing defects but also increasing the T$_C$ should lead to sharper AFE behavior. \ce{La3NbO7} is part of a broader \ce{\textit{A}$_3$\textit{M}O$_7$} family (\textit{A} = rare-earth and \textit{M} = Nb, Ta) and substitution strategies especially on the \textit{A} site can be used to control the energy landscape. This offers many possibilities in tuning electrical response through alloying or strain. For instance, alloying lanthanum with smaller rare-earth (\textit{e.g.}, \ce{(La,Pr)3NbO7} or \ce{(La,Nd)3NbO7}) could be used to increase $T_C$ while keeping the barrier between the AFE and FE phase reasonable and conserving switchability. Alloying is commonly used in ferroic materials and we expect these approaches to be successful on these new weberite-type AFEs. Switching to other rare-earth elements would also open up the possibility to stabilize a very interesting additional FE phase with space group \textit{Amm}2 which possess a large in-plane polarization. 

Finally, this work is a rare example of the successful discovery of a new ferroic material family through the integration of a first principles data-driven screening approach with experiments. While our screening still has limitations and AFEs with more complex distortion mechanisms (\textit{e.g}., resulting from the condensation of several modes) could be missed, our work establishes a methodology to build on and paves the way towards further computationally-driven design and discovery of AFEs.

\section{Conclusion}
Using a database of phonon band structures, we identified a series of potential AFE materials and especially highlighted the weberite-type \ce{La3NbO7}. Thin-film synthesis followed by TEM and electrical characterization demonstrated that \ce{La3NbO7} is indeed a new AFE. \ce{La3NbO7} is the first switchable Kittel-type AFE as the anti-ferroelectricity emerges from the competition of a polar and anti-polar distortion originating from a single phonon mode. Additionally, \ce{La3NbO7} shows attractive properties such as small volume change during switching as well as large threshold and breakdown fields. Our finding opens a new avenue towards an entire new family of weberite-type \ce{\textit{A}$_3$\textit{M}O$_7$} AFEs with the potential for alloying (\textit{e.g.}, through mixing rare-earth on the \textit{A} site) and tuning the electrical response or transition temperature. Our work is also a rare demonstration of the computationally-driven discovery of a new ferroic material and it paves the way towards further computational materials design efforts to search and develop new AFEs.

\section{Methods}

\subsection{Phonon database screening}
The phonon band structure database used to start this work was developed by Atsushi Togo using phonopy \cite{phonopy1, phonopy2}. We defined oxides as materials with an atomic fraction of oxygen larger than 5\%. The materials in this database are found in both the Inorganic Crystal Structure Database \cite{ICSD} (ICSD) and the Materials Project \cite{MaterialsProject}. The children were then obtained by applying the dynamical equation (see Eq. \ref{eq:dynamical_equation}) as implemented in phonopy. To limit the computational cost, the children were generated for the 696 materials which exhibits an instability at $\Gamma$ larger than -5cm$^{-1}$. This criterion helps ensuring that the unstable mode are due to actual structural instabilities rather than to computational artifacts \cite{Pallikara2022}. Finally, 80 materials were further discarded due to problems during the relaxation workflow. To avoid the creation of large cells, we limit the child size to a $2\times2\times2$ supercell of the parent cell.

\begin{equation}
\label{eq:dynamical_equation}
    \mathbf{u}_{j} = \frac{A}{\sqrt{N_a m_j}} Re(\exp(i \phi) \mathbf{e}_{j} \exp(i \mathbf{q} \cdot \mathbf{r}_{jl})
\end{equation}

\subsection{Computational methods}
All calculations were performed with the Vienna Ab-initio Simulation Package (VASP), a density functional theory (DFT) plane-wave code \cite{DFT1, VASP1, VASP2} based on the projected augmented wave method formalism \cite{PAW1}. The energy cutoff for the plane wave basis was set to 520eV. The f-electrons of the Ln were kept frozen by using the Ln\_3 pseudopotentials provided by VASP (except for La for which the pseudopotential was La). For Nb and O, we used 11 (4p$^6$5s$^1$4d$^4$) and 6 (2s$^2$2p$^4$) valence electrons, respectively. For Pb and Zr, we used 14 (5d$^10$6s$^2$6p$^2$) and 12 (4s$^2$4p$^6$5s$^2$4d$^2$) valence electrons, respectively. The initial search was carried out with the Perdew-Burke-Ernzerhof (PBE) exchange-correlation functional but we subsequently used their revised version for solids (PBEsol) for all further calculations \cite{PBE, PBEsol}.  We used a reciprocal density of 200 k-points to sample the Brillouin zone. The electronic self-consistent field loop was considered converged once the difference in energy of two consecutive steps was less than 10$^{-6}$eV per atom while the ionic relaxation was stopped once the forces on the atoms were less than 0.01eV/\r{A}. The band gap of \ce{La3NbO7} was evaluated using the hybrid exchange-correlation functional of Heyd-Scuseria-Ernzherof with 20\% of exact exchange \cite{Heyd2003}. In this case, we used the geometry of the PBE-relaxed cell.

The recalculation of the phonon band structure of the parent structure of \ce{La3NbO7} was done using Pheasy \cite{Lin2025}. For the relaxation of the structure, we set the relaxation criterion on the forces to 10$^{-4}$eV/\r{A}, increased the energy cutoff to 600eV, and used a k-point grid of 8$\times$8$\times$8. To derive the interatomic force constant matrix, the forces were evaluated by finite-differences in a 3$\times$3$\times$2 supercell.

The polarization of the FE phase of all the candidates was calculated using the modern theory of polarization with the parent structure as the reference. The energy barriers were calculated using the nudged elastic band (NEB) approach \cite{Henkelman2000}. The initial guesses for the intermediate states were constructed by linear interpolation between the AFE and FE state. In this case, the ionic relaxation threshold was set to 0.02eV/\r{A}.

The AIMD simulations were performed on a 3$\times$3$\times$2 supercell of the parent phase, corresponding to a simulation cell of 792atoms, and using a NVT. The Nos\'e-Hoover model was used to control the temperature. We set the time step to 2fs and let the simulation run for 5ps for T=100K and T=150K and for 10ps for T=200K to T=350K because of the vicinity to a phase transition. The first 2.5ps of the simulation were left out from the averages to allow the system to reach thermal equilibrium. We used the same pseudopotentials as in our DFT calculations but reduced the energy cutoff of the plane wave basis set to 400eV. Due to the large spatial size of the simulation cell, the Brillouin zone was sampled at the $\Gamma$-only.

\subsection{Thin-Film Growth}
All films were grown using pulsed-laser deposition using a KrF excimer laser (Coherent LPX 305, 248 nm, 25 ns pulse) which was focused onto a dense, stoichiometric, ceramic target of the same composition as the films to deposit material to a prepared substrate in an on-axis geometry with a substrate distance of 55 mm. Prior to growth, the \ce{SrTiO3} substrate was prepared by ultrasonicating for five minutes in acetone and then isopropyl alcohol, after which it was attached to an Inconel resistive heater using silver paint. \ce{La3NbO7} growths were performed at a substrate temperature of 740 °C, dynamic oxygen partial pressure of 100 mTorr, laser fluence of 1.2 J/cm$^{2}$, and laser repetition rate of 2 Hz. LSMO growths were performed at a substrate temperature of 740 °C, dynamic oxygen partial pressure of 200 mTorr, laser fluence of 1 J/cm$^{2}$, and laser repetition rate of 2 Hz. 

\subsection{X-Ray Diffraction}
X-ray $\theta$-2$\theta$ and $\phi$ linescans were used to characterize the structure and epitaxy of the films and were carried out using a Rigaku SmartLab (9 kW rotating anode source) diffractometer using copper K$_{\alpha}$ radiation (1.54 Å wavelength). A hybrid, 2-bounce Ge (220) monochromator and 1/2° divergence slit were used in the incident beam path, and diffracted X-rays were collected with a pixelated area detector (HyPix-3000) operating in 0D (point detection) mode. Prior to all scans, the $\omega$, Z, and $\chi$ axes were aligned to a strong substrate diffraction condition (typically the \ce{SrTiO3} 002).

\subsection{Scanning Transmission Electron Microscopy}
HAADF-STEM imaging and 4D-STEM nanobeam diffraction were performed using a double Cs-corrected FEI Titan Themis G3 (scanning) transmission electron microscope operated at 300 kV. HAADF-STEM images were acquired at a convergence semi-angle of 25 mrad, with a collection semi-angle of 48 to 200 mrad. The 4D-STEM nanobeam diffraction was collected using EMPAD \cite{Tate2016}. The convergence semi-angle was approximately 0.5 mrad. The simulation was conducted using abTEM package \cite{Madsen2023}.

\subsection{Capacitor-Based Dielectric and Polarization Measurements}
All device-based electrical measurements were taken from parallel-plate, out-of-plane-oriented capacitor structures, using a blanket LSMO film as the bottom electrode and 25-$\mu$m-diameter circular contacts as the top electrodes. To produce the top electrodes, following growth of LSMO/\ce{La3NbO7}/LSMO heterostructures on \ce{SrTiO3} (001), the samples were spin-coated in photoresist (polymethyl methacrylate) and cured for five minutes on a hot plate at 100 °C. Photolithography was then performed using a shadow mask under ultra-violet (395 nm) light for five seconds. Exposed patterns were then developed in Microposit MF-319 (tetramethyl ammonium hydroxide) developer for 10-15 seconds. Samples were then etched via argon-ion milling using an Intlvac Nanoquest ion mill (500 V beam voltage, 38 mA beam current, 100 V accelerator, 15 °C stage temperature) for 3-4 minutes at 90° incidence to remove the exposed electrode and 30 seconds at 45° incidence to clean the sidewalls of the devices. After milling, the photoresist was removed by ultrasonicating in acetone (for ~10 sec), leaving behind the circular top contacts used for measurements.

Dielectric constant and loss tangent were measured at 100 kHz as a function of DC bias (converted to electric field by dividing by the film thickness) under a maximum AC excitation of 100 kV/cm using an E4990A Impedance Analyzer (Keysight Technologies, Inc.) at room temperature. Five measurements were taken at each point and averaged together to obtain the plotted response. Polarization versus electric-field and current versus electric-field measurements were taken using a Precision Multiferroic Tester (Radiant Technologies, Inc.) Instantaneous current versus electric-field hysteresis loops were taken using a double bipolar triangular waveform with a set amplitude (20 V) and period (0.01-1 ms per bipolar loop, or 1-100 kHz). The measured current is converted to polarization by integrating over the measurement period and dividing by the device area. Low-temperature measurements were conducted down to 12 K in a CRX-VF cryo-free vacuum probe station (LakeShore Cryotronics), using otherwise the same measurement protocols as room-temperature measurements.

Breakdown measurements were taken using a Precision Multiferroic Tester (Radiant Technologies, Inc.) in a similar manner to hysteresis loops but using a monopolar triangular waveform with a frequency of 100 kHz. Starting with a 5V amplitude, the measurements were repeated, increasing the voltage each time by 1 V, until a short or open circuit was measured (distinguished by a sharp increase in current or near-zero current, respectively) along with visible cracking in the device when viewed through a microscope. The electric field at which this breakdown occurred was then recorded and used to generate Weibull statistics for the sample.

\section{Acknowledgments}
This work was intellectually led and primarily supported by the Materials Project, funded by the U.S. Department of Energy under award DE-AC02-05CH11231 (Materials Project program KC23MP). J.S. acknowledges the U.S. Department of Energy, Office of Science, Office of Basic Energy Sciences, under Award Number DE-SC-0012375 for the development of novel polar materials. X.L. acknowledges support from the Rice Advanced Materials Institute (RAMI) at Rice University as a RAMI Postdoctoral Fellow. X.L. and Y.H. acknowledge support from NSF (FUSE-2329111 and CMMI-2239545) and Welch Foundation (C-2065). X.L. and Y.H. acknowledge the Electron Microscopy Center, Rice University. L.W.M. acknowledges that this work was supported by the Air Force Office of Scientific Research under award number FA9550-24-1-0266. Any opinions, findings, and conclusions or recommendations expressed in this material are those of the author(s) and do not necessarily reflect the views of the United States Air Force.

\printbibliography

\clearpage

\begin{center}
  \Huge\bfseries Supplementary Information
\end{center}

\setcounter{section}{0} % reset section counter
\renewcommand{\thesection}{S\arabic{section}} % prefix "S" to section number

\setcounter{table}{0}
\renewcommand{\thetable}{S\arabic{table}}

\setcounter{figure}{0}
\renewcommand{\thefigure}{S\arabic{figure}}

\section*{Supplementary Figures}

\begin{figure}[!h]
    \centering
    \includegraphics[width=0.9\linewidth]{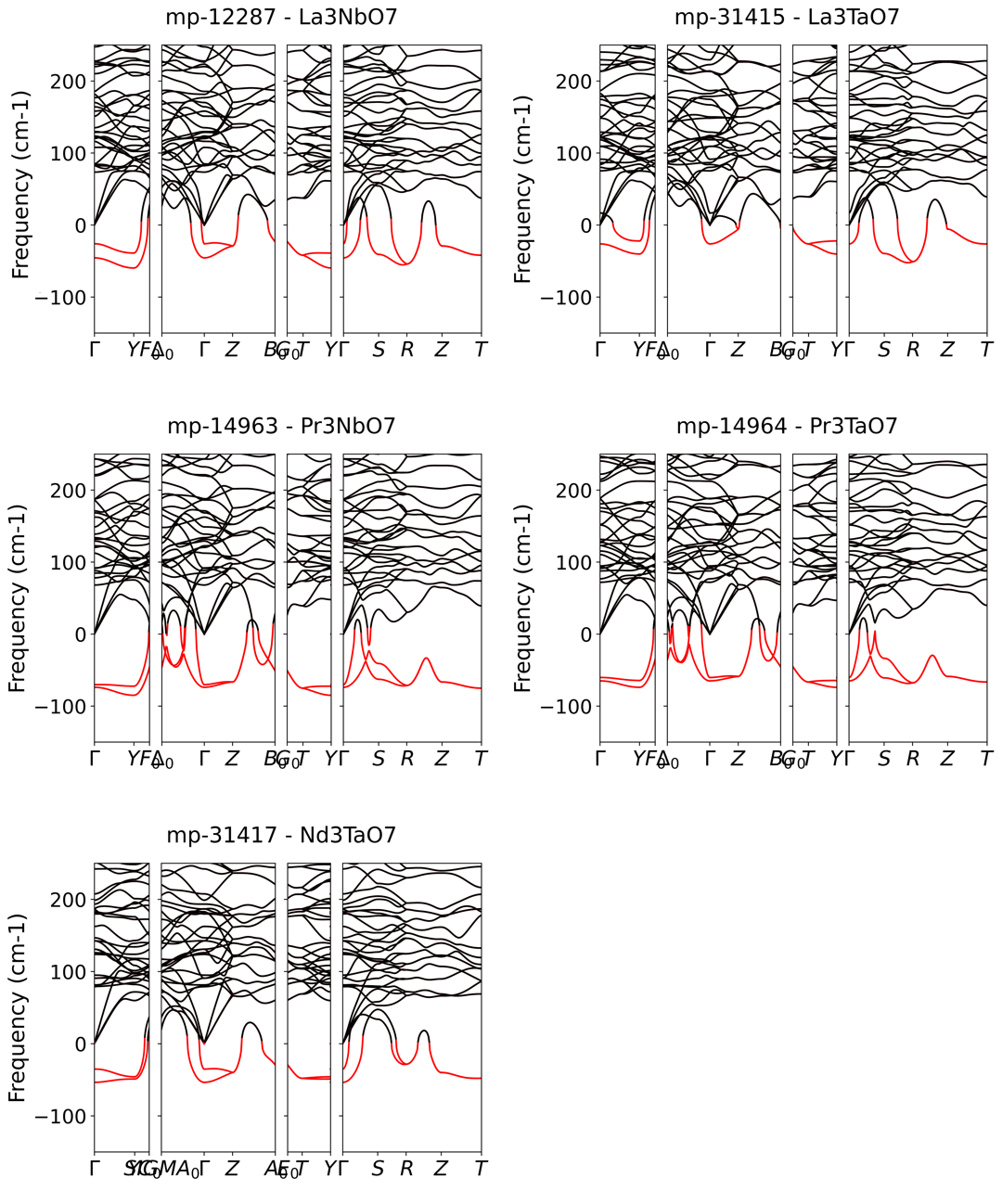}
    \caption{Phonon band structures of the early rare-earths as found in the phonopy phonon database.}
    \label{SM_fig:A3MO7_phonon_1}
\end{figure}

\begin{figure}[!h]
    \centering
    \includegraphics[width=0.9\linewidth]{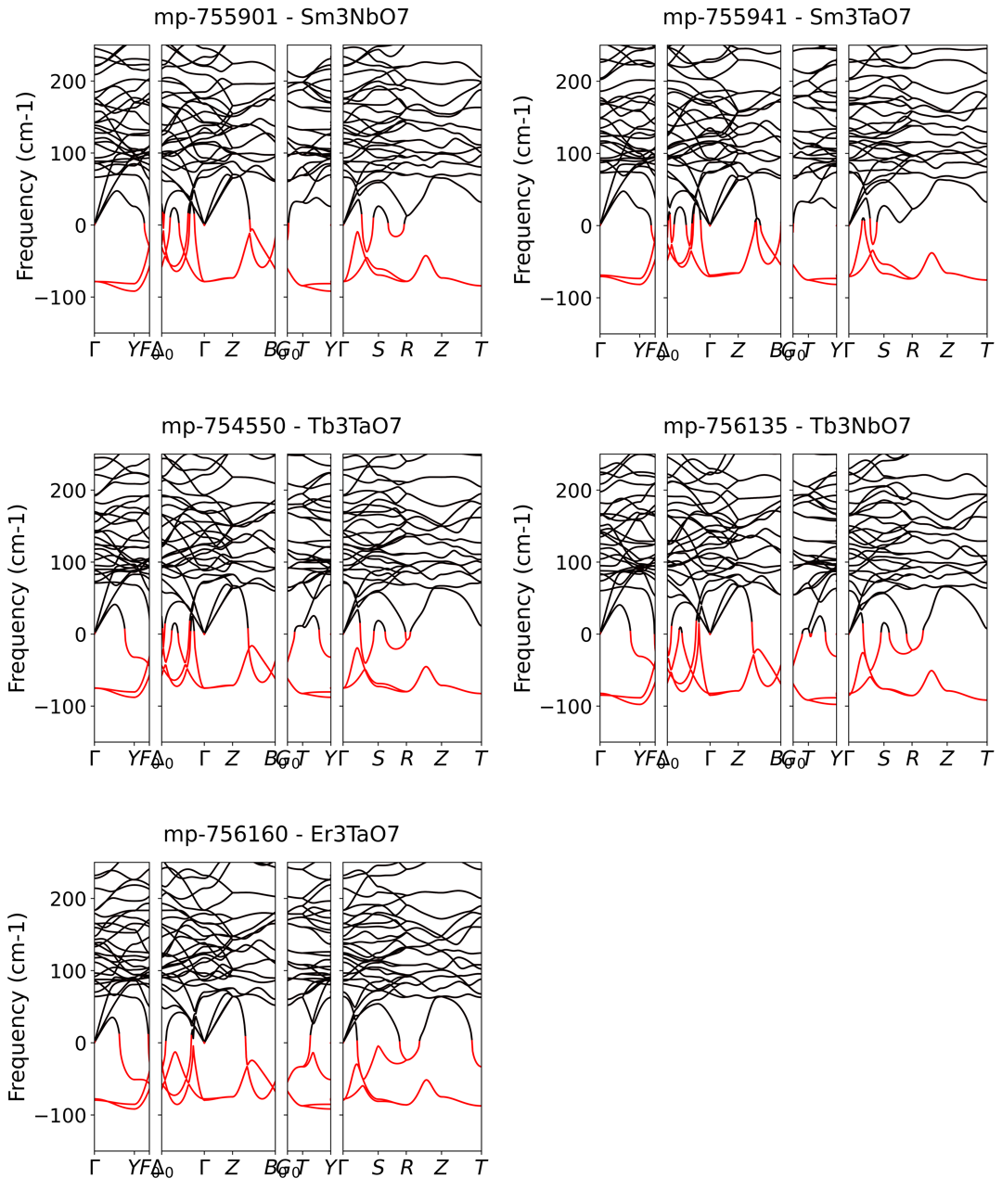}
    \caption{Phonon band structures of the early rare-earths as found in the phonopy phonon database.}
    \label{SM_fig:A3MO7_phonon_2}
\end{figure}

\begin{figure}[!h]
    \centering
    \includegraphics[width=0.75\linewidth]{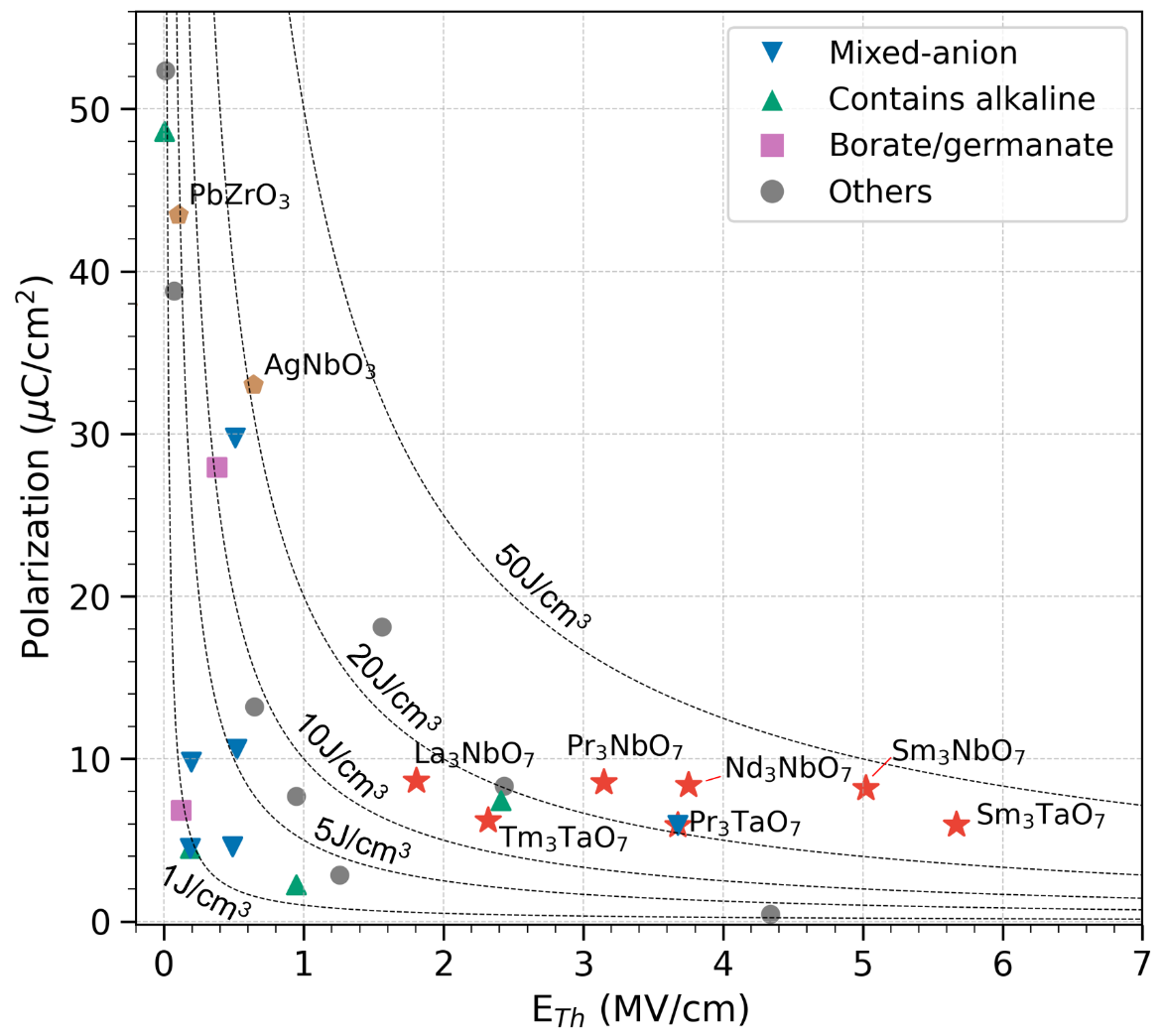}
    \caption{Polarization vs calculated threshold field (E$_{Th}$=$\frac{\Delta E_{AFE-FE}}{P_S \cdot \Omega}$ distribution of the AFE candidates. We also show the energy density isocurves, which to first approximation are given by the product of the threshold field and the polarization. Despite having a relatively small polarization, the large threshold field of the \ce{A3MO7} family leads to large theoretical energy density.}
    \label{SM_fig:P_vs_Eth}
\end{figure}

\begin{figure}[]
    \centering
    \includegraphics[width=0.56\linewidth]{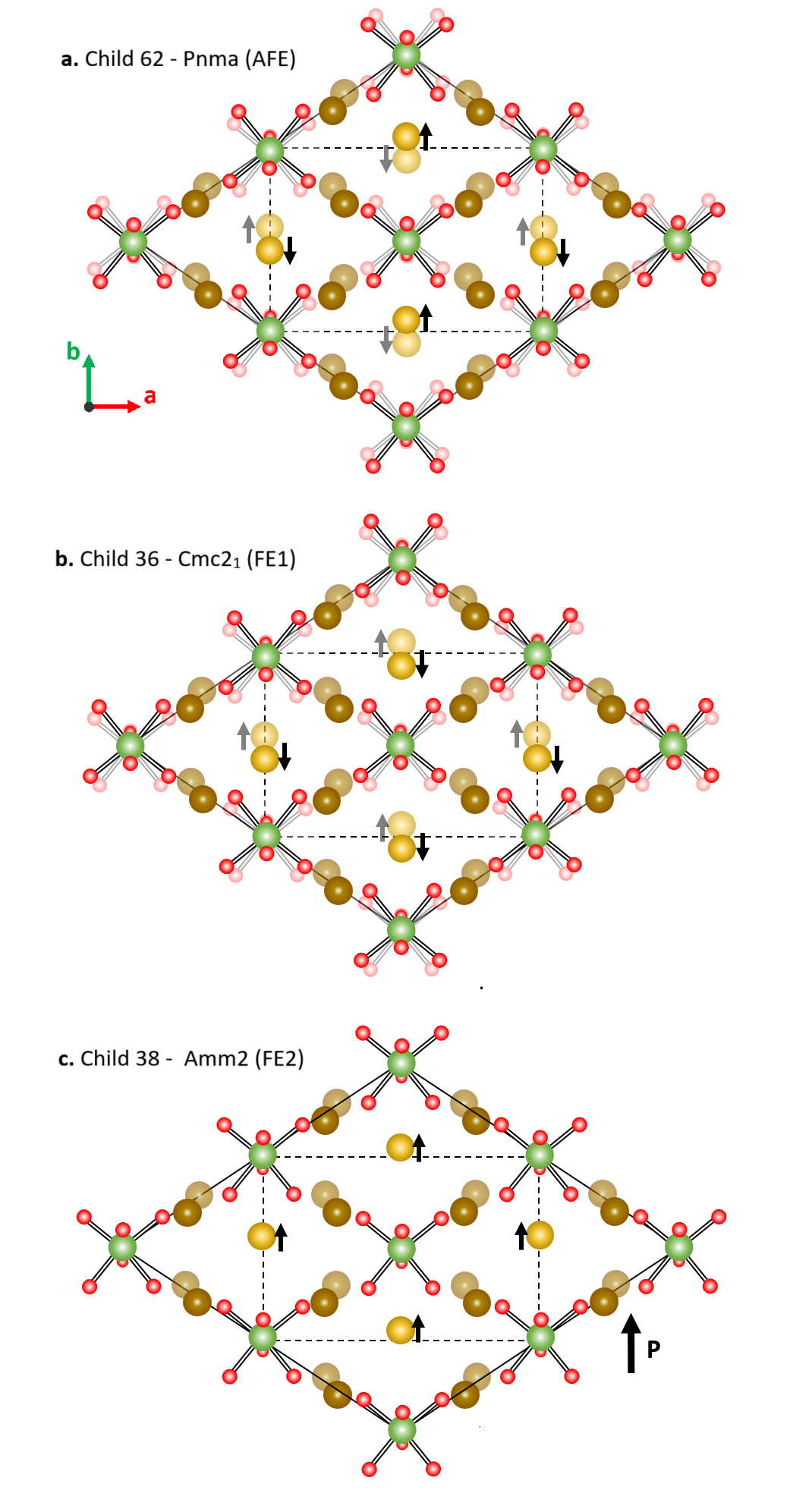}
    \caption{As described in the main text, for rare-earths smaller than La, we found another FE phase in which the planes of 4a A atoms are moving in-phase, leading to a large in-plane polarization. Here, we show the structural differences between the (A) AFE phase Pnma, (B) FE phase Cmc2$_1$ and (C) FE2 phase Amm2, as seen from the \textbf{c} axis. For the FE structure Cmc2$_1$, the polar axis is along \textbf{c} and is mostly associated with the alignment of the Nb inside their oxygen octahedra (not seen here) while the displacement of the 4a A atoms in the [001] plane is antiferroelectric. On the other hand, in the Amm2 structure (FE2), there is a concerted shift of the 4a A atoms along the [010] with virtually no Nb displacements. Due to the large magnitude of these atomic displacement, the resulting polarization in the Amm2 is significantly larger ($\approx$ 30$\mu$C/cm$^2$) than in the Cmc2$_1$ phase ($\approx$ 8$\mu$C/cm$^2$)}
    \label{SM_fig:In-plane_structures}
\end{figure}

\begin{figure}
    \centering
    \includegraphics[width=1.0\linewidth]{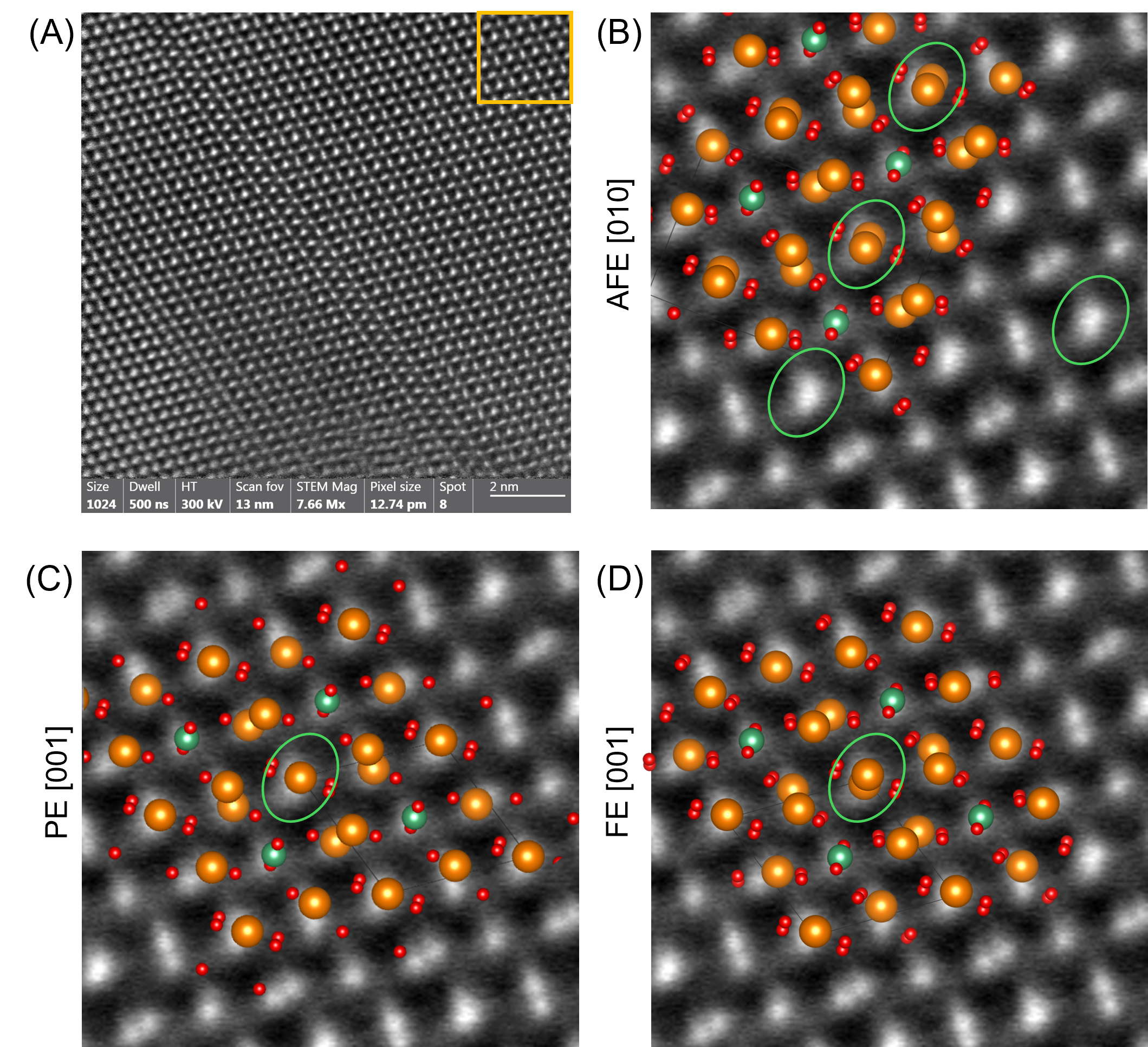}
    \caption{Atomic-resolution HAADF-STEM images taken from \ce{La3NbO7} films. Zooming in from \textbf{a.} low-magnification images, the structure can clearly be overlayed with \textbf{b.} AFE, \textbf{c.} PE, and \textbf{d.} FE phases (assuming [010], [001], and [001] zone axes, respectively). While these structures are nearly indistinguishable in most regards, there are obvious lanthanum distortions in the AFE and FE phases (circled in green) which are not present in the PE phase.}
    \label{fig:tem}
\end{figure}

\begin{figure}
    \centering
    \includegraphics[width=1.0\linewidth]{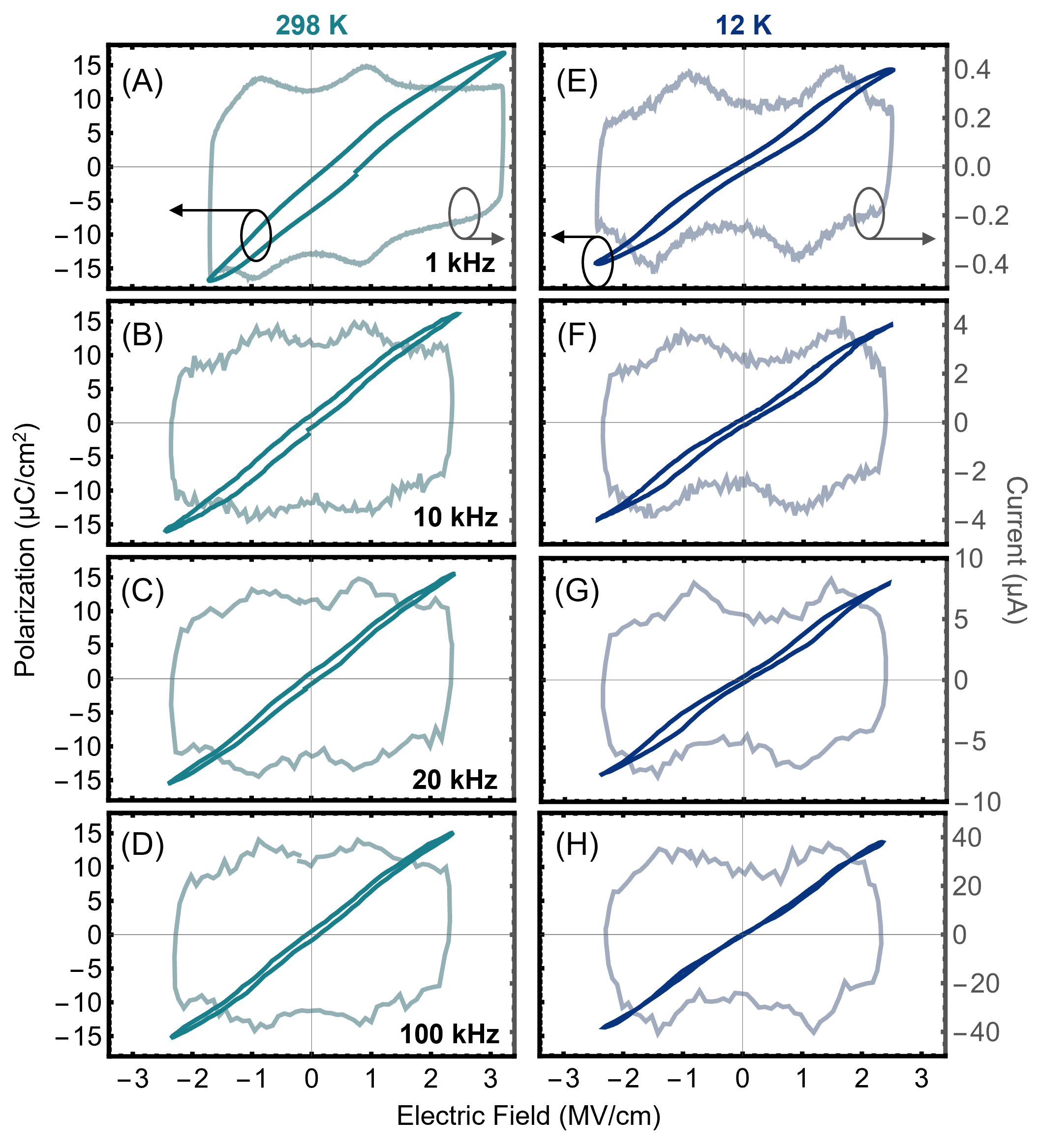}
    \caption{Current and polarization versus electric field loops taken at \textbf{a-d.} room temperature compared to \textbf{e-h.} 12 K, taken at multiple frequencies. All loops show clear indications of antiferroelectric behavior, though the response is expectedly weaker at room temperature.}
    \label{fig:loops}
\end{figure}

\begin{figure}
    \centering
    \includegraphics[width=0.9\linewidth]{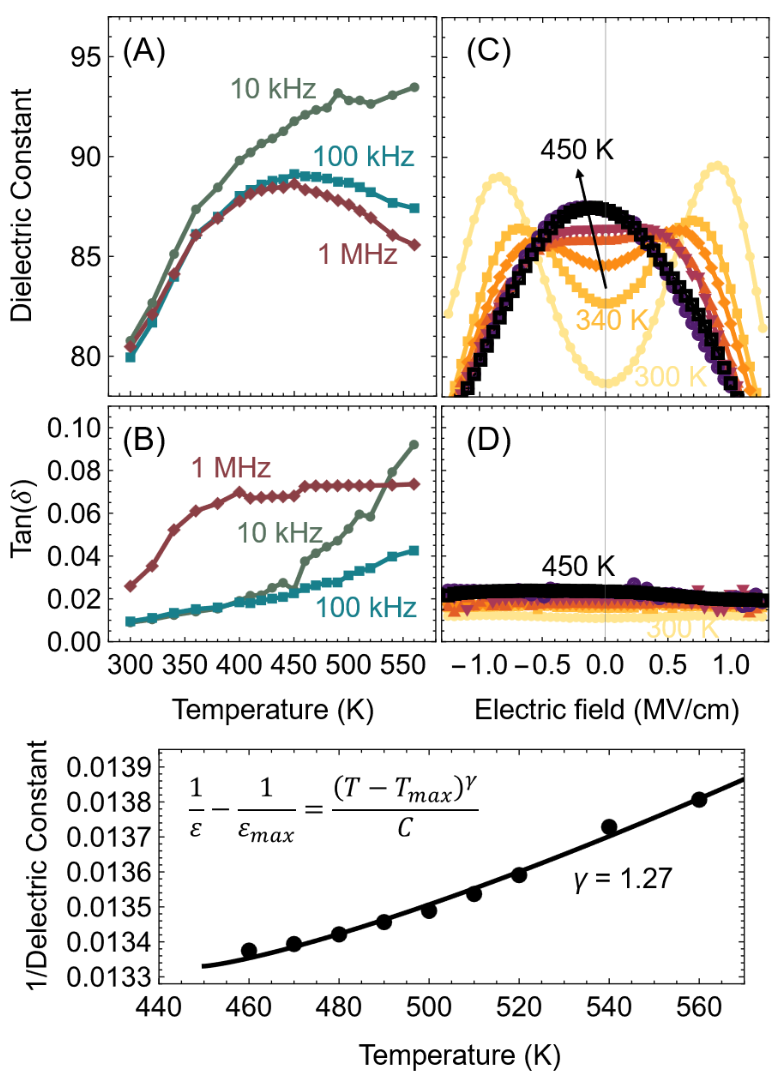}
    \caption{Dielectric measurements as a function of temperature. \textbf{a-b.} dielectric constant and loss at multiple frequencies indicate a transition temperature of ~450 K, while \textbf{c-d.} the dielectric constant versus electric field transitions from antiferroelectric double-peaks to a single peak near this transition temperature. The dielectric response versus temperature was also \textbf{e.} fit to the modified Curie-Weiss law to extract diffusivity.}
    \label{fig:phase-transition}
\end{figure}

\begin{figure}
    \centering
    \includegraphics[width=0.6\linewidth]{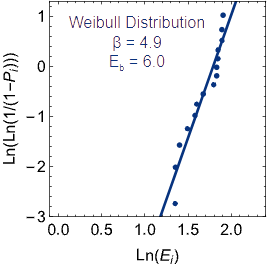}
    \caption{Weibull distribution measurements taken from 15 capacitors used to extract a breakdown strength of 6 MV/cm and Weibull parameter of 4.9.}
    \label{fig:breakdown}
\end{figure}

\begin{figure}
    \centering
    \includegraphics[width=0.95\linewidth]{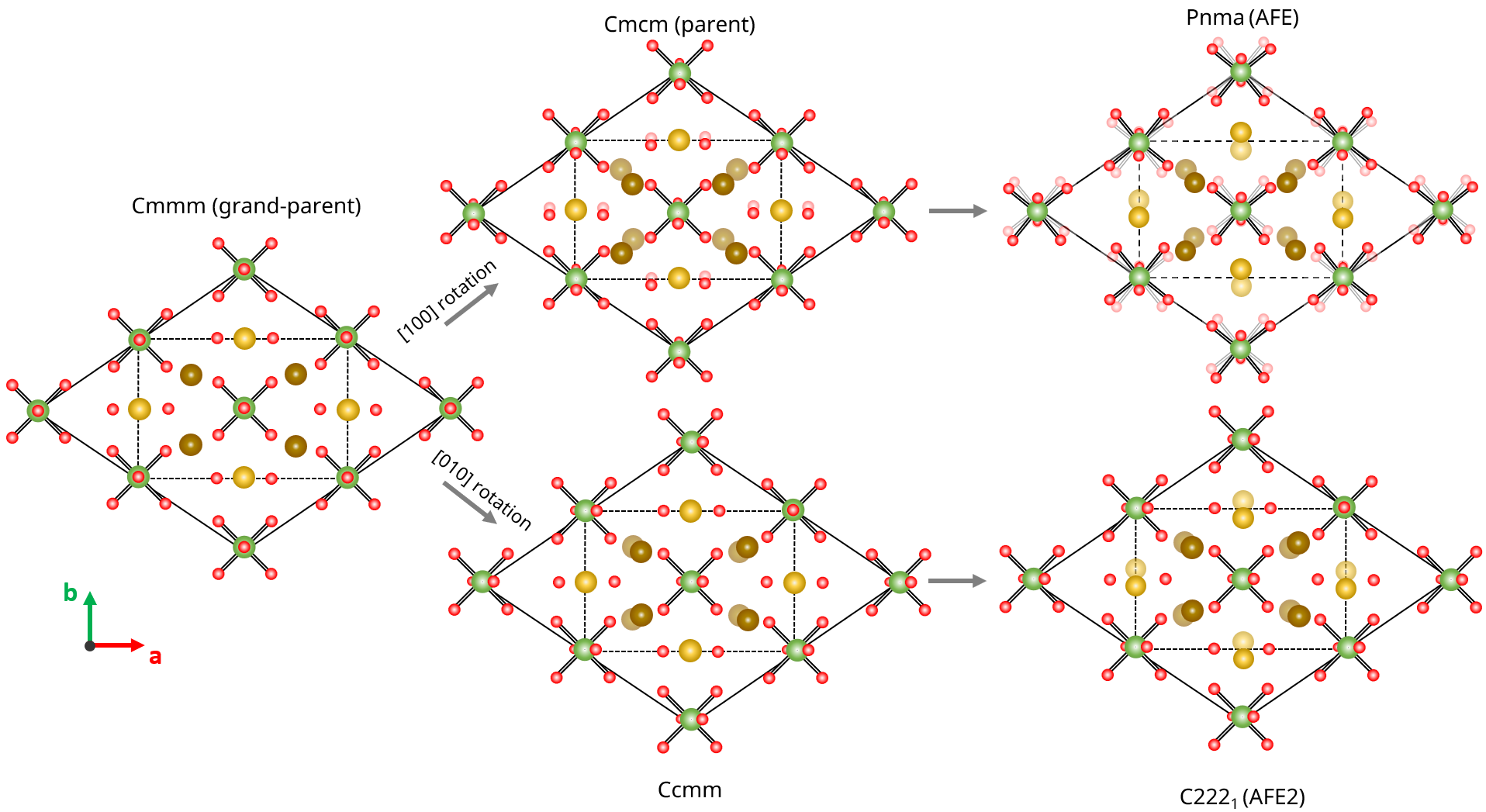}
    \caption{(A) Cmmm "Grand parent" phase in which no octahedra tilt occurs. (B) Cmcm (parent phase found in the phonon database) and (C) Ccmm, a hypothetical phase in which the octahedra are tilted around the [010] axis. (D) the AFE phase Pnma, which originates from an antipolar unstable phonon mode from Cmcm and, (E) the AFE2 phase C222$_1$ phase, which is related to the Ccmm phase.  }
    \label{SM_fig:difference_between_C2221_Pnma}
\end{figure}

\begin{figure}
    \centering
    \includegraphics[width=0.95\linewidth]{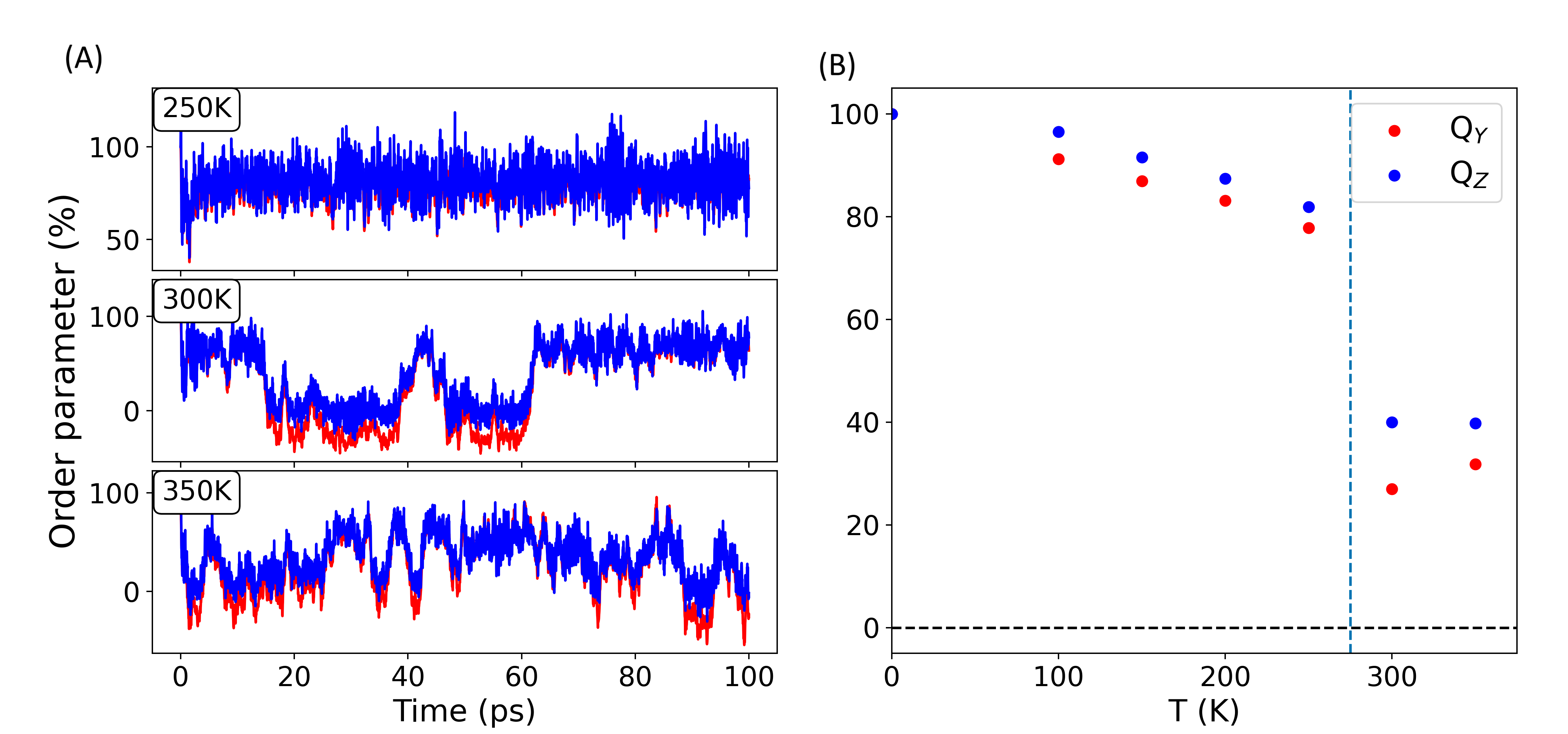}
    \caption{(A) Evolution of the AFE order parameters, $\eta_{y}$ (red) and $\eta_{z}$ (blue), at temperatures close to the transition. Both order parameters are expressed in \%, with respect to their values at 0K, which are $\eta_{y}$=28.48$\mu$C/cm$^2$ and $\eta_{z}$=13.73$\mu$C/cm$^2$. (B) Temperature dependence of $\eta_{y}$ and $\eta_{z}$.}
    \label{SM_fig:AIMD}
\end{figure}

\FloatBarrier

\section*{Supplementary Tables}

% Please add the following required packages to your document preamble:
% \usepackage{graphicx}
\begin{table}[]
\setlength{\tabcolsep}{6pt} % column separation
\renewcommand{\arraystretch}{1.1} % row separation
\centering
\resizebox{\textwidth}{!}{%
\begin{tabular}{lcccccc}
\toprule
formula & AFE-FE & $\Delta$E (meV/at) & V (\AA$^3$/u.c.) & P ($\mu$C/cm$^2$) & E$_{Th, DFT}$ (MV/cm) & E$_{Th, exp}$ (MV/cm) \\
\midrule
\ce{NaNbO3} & Pbcm-R3c  & -0.751* & 60.31 & 37.81 & -0.26 & $\approx$0.1-0.2 \cite{Luo2023} \\
\ce{AgNbO3} & Pmc2$_1$-R3c & +1.614  & 61.13 & 33.04 & 0.64  & $\approx$0.15 \cite{Luo2020}   \\
\ce{PbZrO3} & Pbam-R3c  & +0.394  & 71.72 & 43.47 & 0.1   & $\approx$ 0.3 \cite{Pan2024} \\
\bottomrule
\end{tabular}%
}
\caption{Calculated threshold field of three main representative AFE materials. As previously reported, we find that the calculated ground state of \ce{NaNbO3} is not the AFE phase Pbcm but the FE phase R3c \cite{Amisi2023}. This is consistent with most studies reporting FE-like hysteresis loop in pure \ce{NaNbO3}.}
\label{SM_tab:known_AFEs}
\end{table}

\begin{table}[!h]
\setlength{\tabcolsep}{6pt} % column separation
\renewcommand{\arraystretch}{1.1} % row separation
\centering
\resizebox{\textwidth}{!}{%
\begin{tabular}{lccccc}
\toprule
formula      & $\Delta$E$_{parent}$ (meV/at) & $\Delta$E$_{AFE-FE}$ (meV/at) & $F_{SW}$ (MV/cm)      & P ($\mu$C/cm$^2$) & Eg (eV) \\
\midrule
\ce{NaNbO3}       & -6.19                          & 0.0                            & 0.001            & 48.6                  & 2.0     \\
\ce{SrTaNO2}      & -8.19                          & 0.05                           & 0.011            & 52.4                  & 1.6     \\
\ce{SrTa2Bi2O9}   & -11.39                         & 0.23                           & 0.070            & 38.8                  & 2.19    \\
\ce{Ba3Ti3(BO6)2} & -1.3                           & 0.07                           & 0.120            & 6.9                   & 2.5     \\
\ce{Tb3Si2Cl5O6}  & -5.63                          & 0.1                            & 0.188            & 4.5                   & 4.68    \\
\ce{LiCaTa2O6F}   & -11.44                         & 0.15                           & 0.194            & 9.8                   & 3.8     \\
\ce{K3Nb3Ge2O13}  & -6.34                          & 0.93                           & 0.376            & 28.0                  & 3.0     \\
\ce{Dy3Si2Cl5O6}  & -1.69                          & 0.28                           & 0.489            & 4.6                   & 4.66    \\
\ce{NbOF3}        & -4.37                          & 1.44                           & 0.508            & 29.8                  & 3.83    \\
\ce{NaCaNb2O6F}   & -1.57                          & 0.45                           & 0.520            & 10.6                  & 2.95    \\
\ce{BaTi2O5}      & -1.91                          & 0.67                           & 0.646            & 13.2                  & 2.1     \\
\ce{LiSiBiO4}     & -0.76                          & 0.17                           & 0.945            & 2.3                   & 3.23    \\
\ce{Cd2As2O7}     & -3.05                          & 0.63                           & 0.946            & 7.7                   & 1.64    \\
\ce{La6Ti2S8O5}   & -2.75                          & 0.43                           & 1.256            & 2.9                   & 1.44    \\
\ce{Zn2As2O7}     & -7.24                          & 2.08                           & 1.559            & 18.1                  & 1.95    \\
\ce{La3NbO7}      & -2.87                          & 1.45                           & 1.803            & 8.7                   & 2.79    \\
\ce{Tm3TaO7}      & -11.35                         & 1.13                           & 2.318            & 6.2                   & 3.52    \\
\ce{NaLaMgTeO6}   & -3.85                          & 1.36                           & 2.410            & 7.4                   & 2.99    \\
\ce{TiSO5}        & -5.37                          & 1.67                           & 2.432            & 8.3                   & 2.61    \\
\ce{Pr3NbO7}      & -9.22                          & 2.46                           & 3.148            & 8.6                   & 2.81    \\
\ce{Pr3TaO7}      & -6.63                          & 1.98                           & 3.674            & 5.9                   & 3.51    \\
\ce{Nd3NbO7}      & -10.21                         & 2.82                           & 3.752            & 8.4                   & 2.8     \\
\ce{Ca5Ta4O15}    & -2.94                          & 0.15                           & 4.340            & 0.4                   & 3.4     \\
\ce{Sm3NbO7}      & -12.18                         & 3.55                           & 5.019            & 8.2                   & 2.78    \\
\ce{Sm3TaO7}      & -9.18                          & 2.9                            & 5.670            & 5.9                   & 3.5     \\
\ce{CsLiSO4}      & -3.77                          & 1.05                           & 11.691           & 0.9                   & 5.49    \\
\ce{SiO2}         & -12.5                          & 3.48                           & 555.950          & 0.1                   & 5.54   \\
\bottomrule
\end{tabular}%
}
\caption{List of all the antiferroelectric candidates identified in our search and recalculated with PBEsol. We classify the materials by their switching field.}
\label{SM_tab:AFE_candidates}
\end{table}

\begin{table}[!h]
\centering
\small
\setlength{\tabcolsep}{6pt} % column separation
\renewcommand{\arraystretch}{1.2} % row separation
\begin{adjustbox}{max width=\textwidth}
\begin{tabular}{
    l 
    l 
    S[table-format=2.3]
    S[table-format=2.3]
    S[table-format=2.3]
    S[table-format=2.3]
    S[table-format=2.3]
    S[table-format=2.3]
    S[table-format=2.3] 
}
\toprule
\textbf{Origin} & \textbf{Space Group} & \textbf{La} & \textbf{Pr} & \textbf{Nd} & \textbf{Sm} & \textbf{Gd} & \textbf{Tb} & \textbf{Ho} \\
\midrule
$\Delta_0$ & Cc (9)                                                    & \textbf{-2.741 }& \textbf{-8.989 }& \textbf{-10.365} & -11.897 & -14.277 & -14.831 & -17.179 \\
Y          & Pnma (62)                                                 & -2.691 & -8.952 & -9.989  & -11.853 & -14.186 & -15.854 & -17.082 \\
\cite{Allpress1979, Doi2009, Chesnaud2015, Chen2018} & C222$_1$ (20)   & 2.011  & -4.294 & -8.586  & \textbf{-16.286} & \textbf{-23.722} & \textbf{-26.621} & \textbf{-32.006} \\
$\Gamma$   & Cmc2$_1$ (36)                                             & -1.241 & -6.5   & -7.137  & -8.333  & -9.868  & -10.164 & -11.743 \\
$\Gamma$   & Amm2 (38)                                                 & 0.014  & -4.083 & -4.666  & -5.622  & -6.818  & -7.000  & -8.217  \\
B$_0$      & Pm (6)                                                    & 5.985  & -0.457 & -1.431  & -3.287  & -10.228 & -5.728  & -12.588 \\
F$_0$      & Pm (6)                                                    & -0.7   & 0.8    & -0.12   & -1.839  & -8.592  & -4.035  & -10.71  \\
Y          & Pmmn (59)                                                 & -0.016 & -4.74  & -5.546  & -6.922  & -8.614  & -9.021  & -10.704 \\
$\Delta_0$ & Pm (6)                                                    & 6.057  & 0.8    & -0.12   & -1.839  & -8.594  & -4.035  & -10.71  \\
T          & Cm (8)                                                    & -1.02  & -5.907 & -6.751  &       {---} &       {---} &       {---} &  {---} \\
T          & Aba2 (44)                                                 & {---}  & {---}  & {---}   & -8.271  & -10.187 & -7.912  & -9.273  \\
G$_0$      & Pm (6)                                                    & 5.992  & -0.456 & -1.425  & -3.285  & -10.224 & -5.721  & {---}   \\
Z          & Cm (8)                                                    & -0.03  & -4.683 & -5.299  & {---}   & {---}   & {---}   & {---}   \\
R          & Cm (8)                                                    & -0.053 & -0.052 & -0.059  & -0.007  & -0.009  & -0.011  & -0.048  \\
S          & P21/c (14)                                                & {---}  & -0.037 & -0.047  & -0.058  & -3.311  &         &         \\
Z          & Amm2 (38)                                                 & {---}  & {---}  & {---}   & -6.364  & -7.682  & -3.77   & -5.048  \\
\bottomrule
\end{tabular}
\end{adjustbox}
\caption{Energy stability with respect to the parent phase (in meV/at) of the different children and for a few representative rare-earths of the M=Nb series. Blue shading highlights the lowest energy values per row (most stable).}
\label{SM_tab:energy_landscape}
\end{table}

\begin{table}[!h]
\centering
\small
\setlength{\tabcolsep}{6pt} % column separation
\renewcommand{\arraystretch}{1.2} % row separation
\begin{threeparttable}
\begin{adjustbox}{max width=\textwidth}
\begin{tabular}{
    l
    l
    S[table-format=2.3,table-text-alignment=right]
    S[table-format=2.3,table-text-alignment=right]
    S[table-format=2.3,table-text-alignment=right]
    S[table-format=2.3,table-text-alignment=right]
    S[table-format=2.3,table-text-alignment=right]
    S[table-format=2.3,table-text-alignment=right]
}
\toprule
\textbf{Origin} & \textbf{Space group} & \textbf{La} & \textbf{Pr} & \textbf{Nd} & \textbf{Sm} & \textbf{Tb} & \textbf{Er} \\
\midrule
$\Delta_0$  & Cc (9)        & \textbf{-0.898} & \textbf{-6.460} & \textbf{-9.071}          & -8.920  & -11.443 & -13.793 \\
Y           & Pnma (62)     & -0.879         & -6.432         & -7.515          & -8.878  & -12.439 & -15.394 \\
\cite{Allpress1979, Doi2009, Chesnaud2015, Chen2018} & C222$_1$ (20) & 3.205          & -1.955         & -2.779        & \textbf{-13.284} & \textbf{-23.098} & \textbf{-31.008} \\
$\Gamma$    & Cmc2$_1$ (36) & -0.021         & -4.454         & -5.192        & -6.021  & -7.609  & -9.088  \\
$\Gamma$    & Amm2 (38)     & 0.005          & -2.815         & -3.472        & -4.022  & -5.134  & -6.143  \\
G$_0$       & C2 (5)        & -0.130         & {---}            & {---}           & {---}     & {---}     & {---}     \\
T           & Cm (8)        & -0.045         & -4.264         & -5.173        & {---}     & {---}     & {---}     \\
T           & Imm2 (44)     & {---}            & {---}            & {---}           & -6.211  & -8.204  & -10.032 \\
Y           & Pmmn (59)     & -0.029         & -3.343         & {---}           & -5.084  & -6.818  & -8.395  \\
R           & P$\bar{1}$ (2)      & -0.024         & -0.071         & {---}           & {---}     & {---}     & {---}     \\
Z           & Cm (8)        & -0.018         & -3.229         & -3.909        & {---}     & {---}     & {---}     \\
Z           & Amm2 (38)     & {---}            & {---}            & {---}           & -4.576  & -5.868  & -7.066  \\
S           & P2$_1$/m (11) & 0.010          & -0.042         & {---}           & -0.027  & -0.031  & 0.005   \\
$\Delta_0$  & Pm (6)        & 5.937          & 1.015          & {---}           & -0.123  & -1.932  & -3.250  \\
\bottomrule
\end{tabular}
\end{adjustbox}
\end{threeparttable}
\caption{Energy stability with respect to the parent phase (in meV/at) of the different children and for a few representative rare-earths of the M=Ta series. The parent structure of \ce{Nd3TaO7} found in the database belongs to the space group Ccmm, which differs from the other parent (space group Cmcm) as a result of the oxygen octahedra tilt of the \ce{TaO6} chain being along the [010] axis instead than along the [100] axis. The difference between the two octahedra tilt system is discussed in Fig. \ref{SM_fig:difference_between_C2221_Pnma}. The energy of the phases reported here were obtained by starting from the children and parent (Cmcm) of \ce{Pr3TaO7} and substituting Pr for Nd. }
\label{SM_tab:energy_landscape_Ta}
\end{table}

\begin{table}[]
\centering
\small
\setlength{\tabcolsep}{6pt} % column separation
\renewcommand{\arraystretch}{1.2} % row separation
\begin{tabular}{lcccccc}
\toprule
\textbf{Reference}                               & \textbf{La}                    & \textbf{Nd}                   &  \textbf{Eu}            & \textbf{Gd}        & \textbf{Dy}            & \textbf{Lu}            \\ 
\midrule
Allpress et al. \cite{Allpress1979}     & Cmcm                  & Cmcm                 & ---            & C222$_1$  & ---           & ---           \\
Rossel et al. \cite{Rossell1979}        & Cmcm                  & Cmcm                 & ---            &           & ---           & ---           \\
Kahn-Harari et al. \cite{Kahn1995}      & Pnma                  & ---                  & ---            &           & ---           & ---           \\
Doi et al. \cite{Doi2009}               & Pnma                  & Pnma                 & C222$_1$       & C222$_1$  & Fm$\bar{3}$m  & Fm$\bar{3}$m  \\
Cai et al. \cite{Cai2010}               & ---                   & ---                  & ---            & Cmcm/Cm2m & ---           & ---           \\
Cai et al. \cite{Cai2011}               & Pmcn\tnote{*}$^\ast$  & Pmcn\tnote{*}$^\ast$ & ---            &           & ---           & Fm$\bar{3}$m  \\
Chesnaud et al. \cite{Chesnaud2015}     & Pnma                  & ---                  & ---            & C222$_1$  & ---           & ---           \\
Chen et al. \cite{Chen2018}             & Pnma                  & Pnma                 & Pnma           & C222$_1$  & Fm$\bar{3}$m  & ---           \\ 
Chen et al. \cite{Chen2021}             & Pnma                  & Pnma                 & C222$_1$       & C222$_1$  & ---           & ---           \\
Li et al. \cite{Li2025}                 & ---                   & ---                  & ---            & Ccmm      & ---           & ---           \\ 
\bottomrule
\end{tabular}%
\begin{tablenotes} \footnotesize
\item[*] $^\ast$  Pmcn is a non-standard setting of the space group Pnma.
\end{tablenotes}
\caption{Summary of the experimental reports for six representative rare-earths (La, Nd, Eu, Gd, Dy and Lu).}
\label{SM_tab:A3NbO7_experimental_structure_summary}
\end{table}

% Please add the following required packages to your document preamble:
% \usepackage{graphicx}
\begin{table}[]
\centering
\setlength{\tabcolsep}{6pt} % column separation
\renewcommand{\arraystretch}{1.2}
\begin{tabular}{lcccc}
\toprule
Phase              & a (\AA)   &  b (\AA)  & c (\AA)  & V (\AA)   \\
\midrule
Cmcm (parent)       & 11.1094   &  7.49983  & 7.76517  & 646.98    \\
Pnma (AFE)          & 11.0879   &  7.62635  & 7.73722  & 654.26    \\
Cmc2$_1$ (FE)       & 11.1084   &  7.59857  & 7.73457  & 652.86    \\
\midrule
$\Delta$AFE-FE (\%)  &  0.185   &  -0.364   &  -0.034  & -0.214  \\
\bottomrule
\end{tabular}%
\caption{DFT calculated lattice parameters for \ce{La3NbO7}}
\label{SM_tab:DFT_lattice_parameters}
\end{table}

\FloatBarrier

\section*{Supplementary Notes}

\renewcommand{\thesubsection}{\arabic{subsection}.}
% optional – make the label run into the title (like “1. Title”)

\subsection{Clarification about the ground state of \ce{La3NbO7}}
\label{sm_note:ground_state}
In Table \ref{SM_tab:energy_landscape}, the ground state for the early rare-earth is not the AFE Pnma but the polar phase Cm which obtained from the condensation of a $\Delta_0$ instability. The structural similarity between these two phases makes them almost indistinguishable. After relaxation, the largest ionic displacements between the two phases is only 0.012\r{A}, 0.008\r{A} and 0.018\r{A} for the La, Nb and O atoms, respectively. These differences further decrease if the ionic relaxation criterion is tightened. 

\subsection{Atomic resolution TEM}\label{SM_note:atomic_resolution_TEM}
\label{sm_note:tem}
In addition to TEM-based microdiffraction, atomic-resolution STEM was performed to acquire a real-space picture of the \ce{La3NbO7} film structure, in particular to resolve key differences in the predicted AFE and PE structures. The most distinguishable real-space feature that separates these two structures is a lanthanum distortion in the AFE phase along the [001]/[00$\overline{1}$], which can be viewed along the [010] (Fig. \ref{fig:tem}). Because this distortion alternates between [001] and [00$\overline{1}$], the resulting averaged image down a column of atoms is expected to be a dumbbell shape rather than clearly isolated lanthanum atoms. STEM measurements taken along the [010] of \ce{La3NbO7} films do appear to show faint elongation from these lanthanum sites (circled in green, Fig. \ref{fig:tem} b), though the distortion is close to the resolution limit of the measurement and therefore cannot be unequivocally distinguished over the undistorted version in the PE phase (Fig. \ref{fig:tem} c).

\subsection{Antiferroelectric Polarization and Current Hysteresis Loops}
\label{sm_note:hysteresis}
In addition to low-temperature polarization and current loops shown in the main text, similar measurements were also taken at room temperature and at several difference frequencies (Fig. \ref{fig:loops}). While the \ce{La3NbO7} does still appear to be antiferroelectric at room temperature (based on the four current peaks visible at all measured frequencies, Fig. \ref{fig:loops} a-d), the polarization hysteresis becomes more smeared out, making the characteristic double hysteresis nearly indistinguishable. This is not surprising given the already weak antiferroelectric hysteresis combined with the relative proximity to the phase transition compared to low-temperature measurements.

\subsection{Experimental Phase Transition}\label{SM_note:experimental_phase_transition}
\label{sm_note:exp_phase_transition}
Dielectric measurements were also performed as a function of temperature to characterize the phase transition in \ce{La3NbO7}, where a peak in dielectric response corresponds to a phase transition. Such a peak can clearly be observed in \ce{La3NbO7} around 450 K (Fig. \ref{fig:phase-transition} a), while dielectric loss remains relatively low (Fig. \ref{fig:phase-transition} b), suggesting the response is not an artifact of leakage. Additionally, the characteristic double peaks in dielectric constant versus DC electric field measurements clearly merge into a single peak as temperature increases (Fig. \ref{fig:phase-transition} c-d), as expected for an antiferroelectric going through a high-temperature phase transition. To understand the nature of this phase transition, the dielectric response versus temperature was fit to the modified Curie-Weiss law, and a diffusivity of 1.27 was extracted, indicating the transition is relatively ideal.

\subsection{\ce{La3NbO7} Breakdown Performance}\label{SM_note:La3NbO7_breakdown}
\label{sm_note:breakdown}
To characterize the breakdown performance of \ce{La3NbO7}, Weibull statistics were taken from 15 difference devices measured until breakdown. The resulting breakdown probability was plotted against electric field (Fig. \ref{fig:breakdown}) and fit to the Weibull distribution function:

\begin{equation}
\label{eq:breakdown_field}
    P(E_i) = 1-exp[-(E_i/E_b)^\beta]
\end{equation}

where $P(E_i)$ is the probability of breakdown at electric field $E_i$, $E_b$ is the characteristic breakdown field at which $P(E_i)=$63.2\%, and $\beta$ is the Weibull parameter, which characterizes the distribution of breakdown fields. From these measurements, a large breakdown field of 6 MV/cm was extracted from the \ce{La3NbO7} films.

\subsection{Transition dynamics}\label{SM_note:transition_dynamics}
\label{sm_note:transition_dynamics}
We investigated the transition dynamics of \ce{La3NbO7} using AIMD. The effect of thermal disorder can have dramatic consequence on the polar order and AIMD simulations helps predicting the effect of temperature on the materials property. Nonetheless, it should be emphasized that a precise estimation of the phase transition temperature using this approach is difficult because large cell sizes are required \cite{Zhang2023}. In the case of FE materials, a natural choice for the order parameter is the polarization but it not suited for AFEs and we instead define the anti-polar order parameter $\eta_{\alpha}$:

\begin{equation}
    \eta_{\alpha} = \sum_{N} P_{N,\alpha} (-1)^N
\end{equation}

As the order parameters we monitor during the AIMD simulations. The sum runs over the different sublattices N, and their corresponding polarization $P_{N,\alpha}$ is calculated using the Born effective charges: $P_{N,\alpha} = \frac{e}{\Omega} \sum_{i} Z^*_{i, \alpha \beta} \Delta r_{i,\beta}$. The evolution of $\eta_{y}$ and $\eta_{z}$ over time and for three different temperatures, 250K, 300K and 350K, is shown in Fig. \ref{SM_fig:AIMD} (A). The value of 100\% corresponds to the order parameter of the DFT-relaxed AFE structure. $\eta_{y}$ is mostly associated to the in-plane anti-polar displacements of the La atoms while the largest contributor to $\eta_{z}$ is the out-of-plane displacements of the Nb atoms. At 250K, both $\eta_{y}$ and $\eta_{z}$ are fluctuating over time around a value of 75\%, indicating that we are still below the transition temperature. At 300K, we start observing abrupt changes with $\eta_{y}$ and $\eta_{z}$ oscillating between positive and negative values. Interestingly, these oscillations are in phase, indicating that the atomic motions of Nb and La are somewhat correlated. Finally, at T=350K, the fluctuations become very fast and highlight the dynamical aspect of the transition. 

\end{document}